\documentclass[twocolumn]{aastex62}
\newcommand{\lumcgs}{erg~s$^{-1}$}

\usepackage{natbib}
\graphicspath{{./}{figures/}}

\received{}
\revised{}
\accepted{}
\submitjournal{ApJ}

\shorttitle{The Most Energetic Flare Star GT Mus}
\shortauthors{Sasaki et al.}
\begin{document}

\title{The RS CVn type star GT Mus shows most energetic X-ray flares throughout the 2010s}

\correspondingauthor{Ryo Sasaki}
\email{sasaki@phys.chuo-u.ac.jp}

\correspondingauthor{Yohko Tsuboi}
\email{tsuboi@phys.chuo-u.ac.jp}

\author[0000-0003-3756-6684]{Ryo Sasaki}
\affiliation{Department of Physics, Chuo University, 1-13-27 Kasuga, Bunkyo-ku, Tokyo 112-8551, Japan}
\affiliation{RIKEN, 2-1 Hirosawa, Wako, Saitama 351-0198, Japan}

\author{Yohko Tsuboi}
\affiliation{Department of Physics, Chuo University, 1-13-27 Kasuga, Bunkyo-ku, Tokyo 112-8551, Japan}

\author{Wataru Iwakiri}
\affiliation{Department of Physics, Chuo University, 1-13-27 Kasuga, Bunkyo-ku, Tokyo 112-8551, Japan}
\affiliation{RIKEN, 2-1 Hirosawa, Wako, Saitama 351-0198, Japan}

\author{Satoshi Nakahira}
\affiliation{Institute of Space and Astronautical Science, Japan Aerospace Exploration Agency, 3-1-1
Yoshinodai, Chuo-ku, Sagamihara, Kanagawa 252-5258, Japan}

\author{Yoshitomo Maeda}
\affiliation
{Institute of Space and Astronautical Science, Japan Aerospace Exploration Agency, 3-1-1
Yoshinodai, Chuo-ku, Sagamihara, Kanagawa 252-5258, Japan}

\author{Keith Gendreau}
\affiliation{X-ray Astrophysics Laboratory, NASA/Goddard Space Flight Center, Greenbelt, MD 20771, USA}

\author{Michael F. Corcoran}
\affiliation{CRESST \& the X-ray Astrophysics Laboratory, NASA/Goddard Space Flight Center, Greenbelt, MD 20771, USA}
\affiliation{Institute for Astrophysics and Computational Science, The Catholic University of America, 200 Hannan Hall, Washington, DC 20064}

\author{Kenji Hamaguchi}
\affiliation{CRESST \& the X-ray Astrophysics Laboratory, NASA/Goddard Space Flight Center, Greenbelt, MD 20771, USA}
\affiliation{Department of Physics, University of Maryland Baltimore County, 1000 Hilltop Circle, Baltimore, MD 21250, USA}

\author{Zaven Arzoumanian}
\affiliation{X-ray Astrophysics Laboratory, NASA/Goddard Space Flight Center, Greenbelt, MD 20771, USA}

\author{Craig B. Markwardt}
\affiliation{X-ray Astrophysics Laboratory, NASA/Goddard Space Flight Center, Greenbelt, MD 20771, USA}

\author[0000-0003-1244-3100]{Teruaki Enoto}
\affil{Extreme Natural Phenomena RIKEN Hakubi Research Team, RIKEN Cluster for Pioneering Research, 2-1 Hirosawa, Wako, Saitama 351-0198, Japan}

\author{Tatsuki Sato}
\affiliation{Department of Physics, Chuo University, 1-13-27 Kasuga, Bunkyo-ku, Tokyo 112-8551, Japan}

\author{Hiroki Kawai}
\affiliation{Department of Physics, Chuo University, 1-13-27 Kasuga, Bunkyo-ku, Tokyo 112-8551, Japan}

\author{Tatehiro Mihara}
\affiliation{RIKEN, 2-1 Hirosawa, Wako, Saitama 351-0198, Japan}

\author{Megumi Shidatsu}
\affiliation{Department of Physics, Ehime University, 2-5, Bunkyocho, Matsuyama, Ehime 790-8577, Japan}

\author{Hitoshi Negoro}
\affiliation{Department of Physics, Nihon University, 1-8 Kanda-Surugadai, Chiyoda-ku, Tokyo 101-8308, Japan}

\author{Motoko Serino}
\affiliation{Department of Physics and Mathematics, Aoyama Gakuin University, 5-10-1 Fuchinobe, Chuo-ku,
Sagamihara, Kanagawa 252-5258, Japan}

%% Note that the \and command from previous versions of AASTeX is now
%% depreciated in this version as it is no longer necessary. AASTeX
%% automatically takes care of all commas and "and"s between authors names.

%% AASTeX 6.2 has the new \collaboration and \nocollaboration commands to
%% provide the collaboration status of a group of authors. These commands
%% can be used either before or after the list of corresponding authors. The
%% argument for \collaboration is the collaboration identifier. Authors are
%% encouraged to surround collaboration identifiers with ()s. The
%% \nocollaboration command takes no argument and exists to indicate that
%% the nearby authors are not part of surrounding collaborations.

%% Mark off the abstract in the ``abstract'' environment.
\begin{abstract}

We report that the RS CVn-type star GT Mus (HR4492, HD~101379+HD~101380) was the most active star in the X-ray sky in the last decade in terms of the scale of recurrent energetic flares. We detected 11 flares from GT Mus in 8 yr of observations with Monitor of All-sky X-ray Image (MAXI) from 2009 August to 2017 August. The detected flare peak luminosities were 1--4~$\times$~10$^{33}$~\lumcgs\ in the 2.0--20.0~keV band for its distance of 109.6~pc. Our timing analysis showed long durations ($\tau_{\rm r} + \tau_{\rm d}$) of 2--6 days with long decay times ($\tau_{\rm d}$) of 1--4 days.
The released energies during the decay phases of the flares in the 0.1--100 keV band ranged {1--11~$\times$~10$^{38}$~erg}, which are at the upper end of the observed stellar flare. The released energies during whole duration time ranged {2--13~$\times$~10$^{38}$~erg} in the same band. We carried out X-ray follow-up observations for one of the 11 flares with Neutron star Interior Composition Explorer (NICER) on 2017 July 18 and found that the flare cooled quasi-statically. On the basis of a quasi-static cooling model, the flare loop length is derived to be 4~$\times$~10$^{12}$~cm (or 60~$R_{\sun}$).
%\sout{This is three times larger than the stellar radius (1.7~$\times$~10$^{12}$~cm) and is not exceptionally  large  for a flare from GT Mus.}
The electron density is derived to be {1}~$\times$~10$^{10}$~cm$^{-3}$, which is consistent with  the typical value of solar and stellar flares (10$^{10-13}$~cm$^{-3}$).
The ratio of the cooling timescales between radiative cooling ($\tau_{\rm rad}$) and conductive cooling ($\tau_{\rm cond}$) is estimated  to be $\tau_{\rm rad}$~$\sim$~0.1~$\tau_{\rm cond}$ from the temperature; thus radiative cooling was dominant in this flare.

\end{abstract}

%% Keywords should appear after the \end{abstract} command.
%% See the online documentation for the full list of available subject
%% keywords and the rules for their use.
\keywords{binaries: close --- stars: flare --- X-rays: stars}

%% From the front matter, we move on to the body of the paper.
%% Sections are demarcated by \section and \subsection, respectively.
%% Observe the use of the LaTeX \label
%% command after the \subsection to give a symbolic KEY to the
%% subsection for cross-referencing in a \ref command.
%% You can use LaTeX's \ref and \label commands to keep track of
%% cross-references to sections, equations, tables, and figures.
%% That way, if you change the order of any elements, LaTeX will
%% automatically renumber them.
%%
%% We recommend that authors also use the natbib \citep
%% and \citet commands to identify citations.  The citations are
%% tied to the reference list via symbolic KEYs. The KEY corresponds
%% to the KEY in the \bibitem in the reference list below.

\section{Introduction} \label{sec:int}
{
Stellar flares are thought to be a result of magnetic reconnection
on a stellar surface \citep[e.g.,][]{Shibata_Yokoyama:1999}. The
process has been actively studied in the case of solar flares, since
we can see the flares directly. For the Sun, we can follow the evolution of X-ray emission from plasma loops, which trace the shape of
magnetic fields, and once an abrupt ignition occurs, we see how plasma starts to fill the loops. As for the flares on stars other than
the Sun, on the other hand, the same process has been inferred from
the time variation of physical parameters.}

{Since the beginning of X-ray astronomy, stellar flares have been observed with many instruments (e.g. Einstein, ROSAT, GINGA, ASCA etc.).
These observations have detected the fast rise
and slow decay in the light curves of stellar flares and detected that flare
temperature peaks before the emission measure ({\it EM}), i.e.,
harder emission peaks before softer emission \citep[e.g.,][]{Tsuboi:1998}, all of which are seen in
the solar flares as well.}

{Through these studies, RS CVn systems and
Young Stellar Objects have been recognized as active flare sources.
%The above studies have been done especially for RS CVn systems and
%Young Stellar Objects, both of which are famous to have large flares.
As for the flares from RS CVn stars, in the last two decades, those from UZ
Lib, HR~1099, $\sigma$ Gem, $\lambda$ And and EI Eri were detected with XMM-Newton
\citep{Pandey2012}, those from HR~1099, II Peg, TZ CrB, XY UMa, and AR Lac were detected
with Chandra \citep{Nordon:2007, Drake:2014, Gong:2016}; and those
from II Peg were detected with Swift \citep{Osten:2007}, for example.
%, HR~1099, $\sigma$ Gem, $\lambda$ And and EI Eri, being observed with
%XMM-Newton (7.2 $\times$ 10$^{35}$--4.2 $\times$ 10$^{36}$ erg)
%\citep{Pandey2012}, five flares (HR~1099, II Peg, TZ CrB, XY UMa and AR
%Lac) observed with Chandra
%6~$\times$~10$^{34}$--1.3~$\times$~10$^{36}$ erg) \citep{Nordon:2007,
%  Drake:2014, Gong:2016} and a II Peg flare observed with Swift
%(6~$\times$~10$^{36}$~erg) \citep{Osten:2007}.
However, most of the studies were done with pointed observations,
where the large flares can be detected only by chance, though
there are some rare cases where the pointed observations started with
a trigger by wide-field monitorings \citep[e.g., a flare from II Peg observed with Swift;][]{Osten:2007}.}

{Detection of large flares have increased thanks to the Monitor of All-sky X-ray
Image \citep[MAXI; e.g.,][]{Tsuboi:2016}.
%The MAXI gives more frequent chances to investigate the largest flares.
%We have been monitoring stars in the X-ray band with the Monitor of All-sky X-ray
%Image (MAXI; Matsuoka et al. 2009 \cite{Matsuoka+09}).
MAXI is an all-sky X-ray monitor that has been operating on the Japanese
Experiment Module (JEM; Kibo) on the International Space Station (ISS)
since 2009 August 15 \citep{Matsuoka:2009}. It observes a large area of the sky once per 92 minute
orbital cycle and makes it possible to search for transients
effectively. }

%Here, we report the results with the gas proportional
%counters (GSC) of MAXI obtained in the first 10-year operation from
%2009 August.

%For the recent studies
%(UZ Lib, HR~1099, $\sigma$ Gem, $\lambda$ And and EI Eri) observed with
%XMM-Newton (7.2 $\times$ 10$^{35}$--4.2 $\times$ 10$^{36}$ erg)
%\citep{Pandey2012}, five flares (HR~1099, II Peg, TZ CrB, XY UMa and AR
%Lac) observed with Chandra
%6~$\times$~10$^{34}$--1.3~$\times$~10$^{36}$ erg) \citep{Nordon:2007,
%  Drake:2014, Gong:2016} and a II Peg flare observed with Swift
%(6~$\times$~10$^{36}$~erg) \citep{Osten:2007

%The largest solar flares
%ever observed released energies of the order of $10^{32}$~erg, whereas
%stellar flares have released up to $10^{38}$~erg
%\citep[e.g.,][]{Tsuboi:2016}.

%Stellar flares are thought to be a resultant of magnetic reconnection
%on a stellar surface.

%However, despite of the analogy, the
%geometries of the reconnected magnetic loops are not fully understood
%in the case of large flares, especially. For example, one proposed
%that the loops are connected between binary components, and the other
%one assumed that they are connected between the star and the
%circumstellar disk.

%In order to investigate the origin of the giant flares,

%So far, the most comprehensive observations of stellar flares have been carried out with the all-sky X-ray monitor MAXI \citep[Monitor of All-sky X-ray Image,][]{Matsuoka:2009} which observed stellar flares with flare energies between 10$^{33}$--10$^{38}$~erg from 2009 to the present (see Section~\ref{subsec:maxi} for details).

{\cite{Tsuboi:2016} analyzed stellar flares detected  in  2 yr of MAXI observations. The observed parameters of all of these MAXI/GSC flares are found to be near the upper range for observed stellar flares (see their Figure 4 and 5), with luminosities of $10^{31-34}$ ergs s$^{-1}$ in the 2--20 keV band, EMs of 10$^{54-57}$ cm$^{-3}$, e-folding times of 1 hr to 1.5 days, and total radiative energies of 10$^{34-39}$ ergs.} They found a universal correlation between the flare duration and peak X-ray luminosity, combining the X-ray flare data of nearby stars and the Sun (their Figure 5).

%{Recently, some stellar flares are observed with XMM-Newton, Chandra and Swift \citep[e.g.][]{Pandey2012,Gong:2016}. Especially, an II Peg flare observed with Swift showed 6 $\times$ 10$^{36}$ erg \citep{Osten:2007}. This is one of the largest X-ray flare observed so far. They are not included in \cite{Tsuboi:2016}. However, they are also consistent with the correlation.} Moreover, they found that the MAXI-detected flares extended the established correlation between the flare-peak emission measure and temperature for solar flares and small stellar flares \citep{Shibata_Yokoyama:1999}.
%{The recent flare works are also consistent with this correlation.} These correlations hold over a broad range of energies, from solar micro flares to large stellar flares. This correlation suggests a common mechanism connecting flare-triggering and cooling processes.

Among the MAXI-detected stellar flare sources, the RS CVn-type star GT Mus showed remarkably energetic flares with energies up to $\sim$10$^{38}$~erg, repeatedly. So far, MAXI has detected flare candidates with the MAXI ``nova-alert system'' \citep{Negoro:2016} designed to detect transients from MAXI all-sky images in real time. The MAXI team reported nine flare candidates to the MAXI mailing list. Among them, they reported three candidates to {\it the astronomer's telegram}~\footnote{\url{http://astronomerstelegram.org/}}  (ATel) \citep{Nakajima:2010, Kaneto:2015, Sasaki:2016}. One of them has already been reported in \cite{Tsuboi:2016}.

The quadruple system GT Mus (HR4492) consists of two binary systems named HD~101379 and HD~101380, located at (R.A., Dec.)(J2000) = (11$^{\rm h}$39$^{\rm m}$29$^{\rm s}$.497, $-$65$\arcdeg$23$\arcmin$52$\arcsec$.0135) at a distance of 109.594~pc \citep{Gaia:2016, Gaia:2018}. The two binaries (HD~101379 and HD~101380) are separated by 0.23~arcsec, which is spatially resolved by speckle methods \citep{McAlister:1990}.

The RS CVn-type single-lined spectroscopic binary HD~101379 \citep{Strassmeier:1988, McAlister:1990} has a G5/8 giant primary with a radius of 16.56~$R_{\sun}$ \citep{Gaia:2016, Gaia:2018}. This binary shows strong CaII H, CaII K, and variable H$\alpha$ emissions \citep{Houk:1975}. Moreover, it shows a periodic photometric variation of 61.4 days, which dominates any other variations of GT Mus. This 61.4 day variation may be attributed to a rotational modulation of one or more starspots on HD~101379 \citep{Murdoch:1995}. These features indicate high magnetic activity, which implies that the flare observed by MAXI may have originated on HD~101379.

The other system, HD~101380, is a binary consisting of an A0 and an A2 main-sequence star \citep{Houk:1975, Collier:1982b}. In the folded V-band GT Mus light curve, a small dip is detected \citep{Murdoch:1995}. It is  interpreted to be due to an eclipse of this binary with a period of 2.75 day. No variations by spots have  ever been  observed. Thus, it is feasible to  speculate that HD~101379 has higher chromospheric activity than HD~101380.

All of the reported MAXI flares from GT Mus  so far have been detected by the MAXI ``nova-alert system'' \citep{Negoro:2016}. However, there is a real potential that some flares have been missed by this automated system.  Given the current small number (23) in the MAXI stellar flare sample \citep{Tsuboi:2016} and the highly active nature of GT Mus, GT Mus provides a good opportunity to  study the physical characteristics of stellar flares and their mechanism.

In this work, we carry out a detailed analysis of the MAXI data (Section~\ref{sec:obs}) of GT Mus to search for X-ray flares.  We successfully detect 11 flares (including the three that have been already reported), all of which show a total released energy of 10$^{38}$ erg or higher, and perform a unified analysis for all of them (Section~\ref{sec:res}). In addition, we also carry out  follow-up X-ray observations with Neutron star Interior Composition Explorer \citep[NICER, see Section~\ref{sec:obs};][]{Gendreau:2016} for one of the flares, perform time-resolved spectroscopy, and give much tighter constraints on the physical characteristics (Section~\ref{sec:res}). We then discuss the cooling process of the flare observed with NICER and also GT Mus flares in general in a broader context (Section~\ref{sec:dis}), before summarizing our result (Section~\ref{sec:sum}).

\section{Observations} \label{sec:obs}
\subsection{MAXI}\label{subsec:maxi}
MAXI \citep{Matsuoka:2009} is an astronomical X-ray observatory mounted on the International Space Station (ISS). In this analysis, we used data from  the Gas Slit Camera \citep{Mihara:2011} only, which is sensitive in the 2--30~keV band. It consists of 12 proportional counters, each of which employs carbon-wire anodes to provide one-dimensional position sensitivity. A pair of counters forms a single camera unit; hence the instrument consists of 6 camera units. The six camera units are assembled into two groups whose field of views (FoVs) are pointed toward the tangential direction of the ISS\ motion along the earth horizon and the zenith direction. The FoVs are 160$\arcdeg$~$\times$~3$\arcdeg$, which corresponds to 2\% of the whole sky. These counters are not operated in the regions with high particle background, such as the South Atlantic Anomaly and at absolute latitudes higher than $\sim$40$\arcdeg$, and the vicinity of the Sun (within $\sim$5~$\arcdeg$).
Hence the Gas Slit Camera has an operating duty ratio of $\sim$40$\%$ and scans about 85$\%$ of the whole sky per orbit of the ISS.

In this work, we used the {\it MAXI on-demand system}~\footnote{\url{http://maxi.riken.jp/mxondem/}} \citep{Nakahira:2013} to obtain images, light curves, and spectra. We extracted  source  photons from a  circular region with a radius of 1$\arcdeg$.5 centered on the GT Mus, the area of which corresponds to the point spread function of the Gas Slit Camera. The background  photons  were extracted from a circular region with a radius of 4$\arcdeg$.0 centered at (R.A., Dec)(J2000) = (11$^{\rm h}$24$^{\rm m}$3$^{\rm s}$.7699, $-$67$\arcdeg$4$\arcmin$42$\arcsec$.939), excluding the source area of radius of 2$\arcdeg$.0 centered at the same position  as the source region. Here the center of the background region was shifted slightly from that of the source region in order to avoid  light leakage  from nearby bright sources (Cen X-3 and V830 Cen).

\subsection{NICER}\label{subsec:nicer}

We carried out follow-up observations of a GT Mus flare (FN~11) with
NICER. NICER is a nonimaging X-ray detector installed on the ISS in
2017 June. X-ray detector of NICER, the X-ray Timing Instrument
\citep[XTI,][]{Prigozhin:2012} consists of 56 co-aligned X-ray
concentrator optics (XRCs) and silicon-drift detectors (SDDs). Each
XRC collects X-ray photons over a large geometric area from a 15
arcmin$^{2}$ area of sky. The XRCs concentrate photons onto the
SDDs. The SDDs have a sensitivity in the 0.2--12~keV band with an
energy resolution of 85~eV at 1~keV. The XTIs provide a large
effective area of $\sim$1,900~cm$^{2}$ at 1.5~keV. In practice, out of
56~XRCs, 52~XRCs are operated in orbit.

The NICER follow-up observation was carried out for FN~11. The ``nova-alert system'' \citep{Negoro:2016} triggered a transient event from GT Mus on 2017 July 17 03:55 UT.
The NICER follow-up observation started on 2017 July 18 17:00 UT, $\sim$1.5 days after the trigger, and ended on 2017 July 21 14:36 UT. During the observation, the count rate decayed from 300 to 140 counts per second in the 0.5--10.0~keV band. After 123~days from the MAXI trigger for FN~11, NICER observed GT Mus again (from 2017 November 18 to 2017 November 20). During the 3 day observation, the count rate was constant at
$\sim$43 counts s$^{-1}$ in the 0.5--10.0~keV band. No significant variability during the observation was detected (see Section~\ref{sec:time_res} for details). Moreover, this is in agreement with the count rate of GT Mus in the XMM-Newton slew survey catalog \citep{Freund:2018}, where {\it Web PIMMS} \footnote{\url{https://heasarc.gsfc.nasa.gov/cgi-bin/Tools/w3pimms/w3pimms.pl}} was employed for the count rate conversion. Because of that, we considered these data as the quiescent state of GT Mus.

With NICER, the spectral and temporal  parameters of stellar flares can be determined with a much higher precision than with MAXI. NICER can swiftly respond to emergent observations because the ISS is in real-time contact for $\sim$70\% of a day, thanks to the Tracking and Data Relay Satellite.

In this work, we used all available GT Mus NICER data (observation IDs of 1100140101--1100140108).  The data were calibrated and screened using the HEASARC's \texttt{HEAsoft} package (v6.25), which includes \texttt{NICERDAS} version 5, with the NICER CALDB version 20181105. We processed the data using the task \texttt{nicerl2}, which generates a list of calibrated, accepted photons excluding  periods of especially high background. By this cleaning, the data of observation ID 1100140105  were totally excluded. We extracted source spectra from the cleaned calibrated events.

We estimated background spectra for each of the extracted source spectra.  The NICER background is produced by  charged particles   in the orbit of the ISS, which depends on magnetic cutoff rigidity and space weather conditions.  In addition, optical loading from the sunlight falling on the detectors also contributes to background contamination. Most of them appear in the energy range below 0.4~keV. To estimate the background, we extracted NICER photon events from more than 970~ksec of NICER blank-sky field observations  that have similar the cutoff rigidity, space weather, and sun-angle conditions  to those during the GT Mus observations. We accumulated spectra for these extracted background events and subtracted these spectra from the GT Mus spectra for spectral analysis. The estimated background rates were $<3$ NICER XTI counts per second for all the GT Mus observations. We conservatively used the 0.5--10~keV energy band in the analysis excluding the lowest and highest energy bands of the SDDs, in order to minimize the effects of the low-energy noise and of large calibration uncertainty in the high-energy band above 10~keV.

\section{Results} \label{sec:res}

\subsection{Flare search with MAXI}

\begin{figure*}[htbp]
    \begin{center}
    \includegraphics[width=16cm]{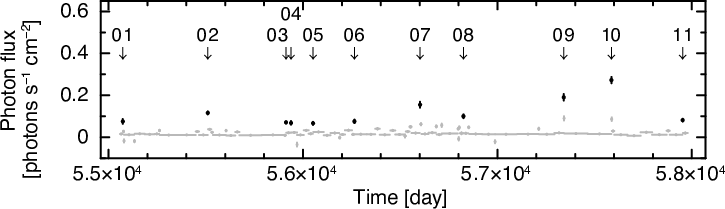}
    \end{center}
    \caption{ MAXI Bayesian block light curve for GT Mus spanning 8 yr (from 2009 August 15 to 2017 August 14). The downward-pointing arrows and black points show the epochs of the detected flares. The numbers above the arrows are the flare IDs, which  correspond to the flare IDs  in Table~\ref{tb:obsevedparameter}.}
    \label{fig:MAXI_bayesian}
\end{figure*}

\begin{center}
	\begin{table*}[htbp]
	%\begin{minipage}
	\caption{Flare epoch, photon flux, and significance of detection with MAXI.}
	%\centering
	\begin{center}
	{\begin{tabular}{lccccccc} \hline
		Flare  ID & MJD$^{a}$ & UT$^{b}$ & Photon flux$^{c}$ & Significance & Detection$^{d}$\\
	     & & & [photons~s$^{-1}$~cm$^{-2}$] & [$\sigma$] & method\\ \hline
	    FN~01 & 55073.691~$\pm$~0.395 & 2009 Aug 30 16:35 & 0.08~$\pm$~0.01 & 10.2 & B\\
		FN~02$^{e}$ & 55510.517~$\pm$~1.016 & 2010 Nov 10 08:34 & 0.12~$\pm$~0.01 & 21.8 & n, B\\
		FN~03 & 55912.883~$\pm$~0.834 & 2011 Dec 17 21:11 & 0.07~$\pm$~0.01 & 19.9 & n, B\\
		FN~04 & 55938.629~$\pm$~0.609 & 2012 Jan 12 15:05 & 0.07~$\pm$~0.01 & 6.0 & n, B\\
		FN~05 & 56051.871~$\pm$~1.349 & 2012 May 04 20:54 & 0.07~$\pm$~0.01 & 8.1 & n, B\\
		FN~06 & 56264.539~$\pm$~1.256 & 2012 Dec 03 12:56 & 0.08~$\pm$~0.01 & 15.0 & n, B\\
		FN~07 & 56602.168~$\pm$~1.096 & 2013 Nov 06 04:01 & 0.15~$\pm$~0.01 & 9.1 & n, B\\
		FN~08 & 56824.965~$\pm$~1.127 & 2014 Jun 16 23:09 & 0.10~$\pm$~0.01 & 9.4 & n, B\\
		FN~09$^{f}$ & 57341.297~$\pm$~0.482 & 2015 Nov 15 07:07 & 0.19~$\pm$~0.02 & 18.4 & n, B\\
		FN~10$^{g}$ & 57585.996~$\pm$~0.450 & 2016 Jul 16 23:54 & 0.27~$\pm$~0.01 & 25.3 & n, B\\
		FN~11 & 57952.516~$\pm$~1.415 & 2017 Jul 18 12:22 & 0.08~$\pm$~0.01 & 19.1 & n, B\\  \hline
	\multicolumn{5}{l}{${a}$ : Detection time of the flare in the adaptive binning (see the text for details).} \\
	\multicolumn{5}{l}{${b}$ : Center time of the observation.} \\
	\multicolumn{5}{l}{${c}$ : The energy band is in the 2.0--10.0~keV band.} \\
	\multicolumn{5}{l}{${d}$ :  Here ``n'' and ``B'' are the nova-alert system and the Bayesian block, respectively.} \\
	\multicolumn{5}{l}{${e}$ : \cite{Nakajima:2010} and \cite{Tsuboi:2016}} \\
	\multicolumn{5}{l}{${f}$ : \cite{Kaneto:2015}} \\
	\multicolumn{5}{l}{${g}$ : \cite{Sasaki:2016}} \\
	\end{tabular} }
	\end{center}
	\label{tb:obsevedparameter}
	%\end{minipage}
	\end{table*}
	\end{center}

We searched for flares from the MAXI GT Mus light curve using data
from 2009 August 15 to 2017 August 14. First, we applied an adaptive
binning with a Bayesian block algorithm \citep{Scargle:2013} to a
one-orbit light curve. Then, we identified statistically significant
variations in the binned light curve with a simple nonparametric
model (Figure~\ref{fig:MAXI_bayesian}), where the false positive rate
(i.e., probability of falsely detecting a change point) was set to
$p_{0}$~=~0.1, which follows that the significance of a change point
is 90\% (=~$1-p_{0}$). {In Figure~\ref{fig:MAXI_bayesian}, the data
points with time bin less than 0.15 day are deleted, because with
such short time bin, the error of the background-subtracted events
cannot be approximated to Gaussian. As flare candidates, the bins with
a photon flux higher than 0.05~photons~s$^{-1}$~cm$^{-2}$ are
selected.} We further filtered out dubious flare candidates using the spatial
significance-checking method employed in \cite{Uzawa:2011} and
\cite{Tsuboi:2016}, in which the threshold significance was set to
5$\sigma$, where $\sigma$ is the standard deviation of the X-ray
counts of the background region in each MAXI image in the
2.0--10.0~keV band, scaled to the source area. We applied this method because the background count rates in the source region are higher than the source count rates in the quiescent state by a factor of 6. Consequently, we found
11 flares. Figure~\ref{fig:MAXI_bayesian} shows the binned light
curve with the 11 flares indicated, and
Table~\ref{tb:obsevedparameter} summarizes the parameters of the
flares, including the 2.0--10.0~keV photon flux (see
Section~\ref{sec:maxi_result} for details) and the detection
significance.

\subsection{MAXI light curves and spectra} \label{sec:maxi_result}

\begin{figure*}[t]
	\begin{minipage}{0.5\hsize}
	\begin{center}
		\includegraphics[width=8cm]{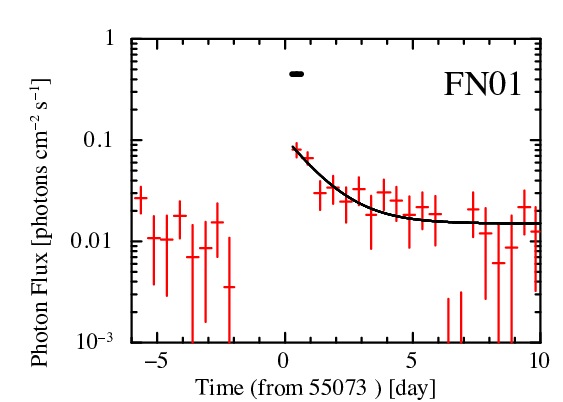}
	\end{center}
	\end{minipage}
	\begin{minipage}{0.5\hsize}
	\begin{center}
		\includegraphics[width=8cm]{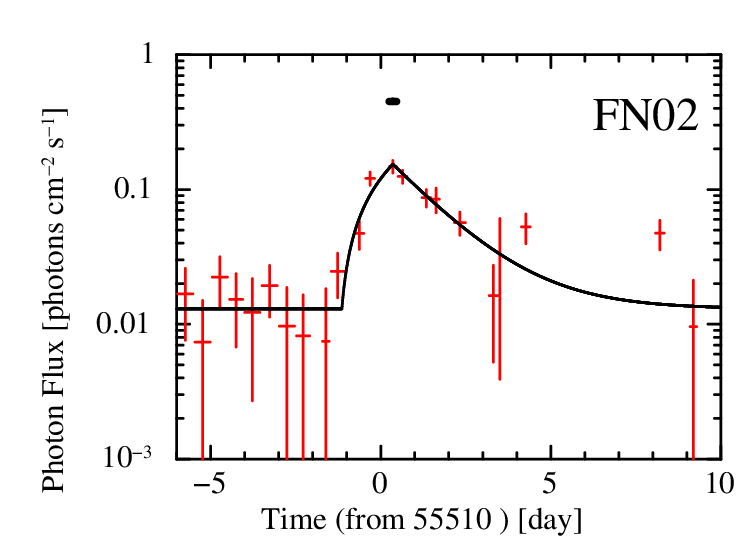}
	\end{center}
	\end{minipage}
	\begin{minipage}{0.5\hsize}
	\begin{center}
		\includegraphics[width=8cm]{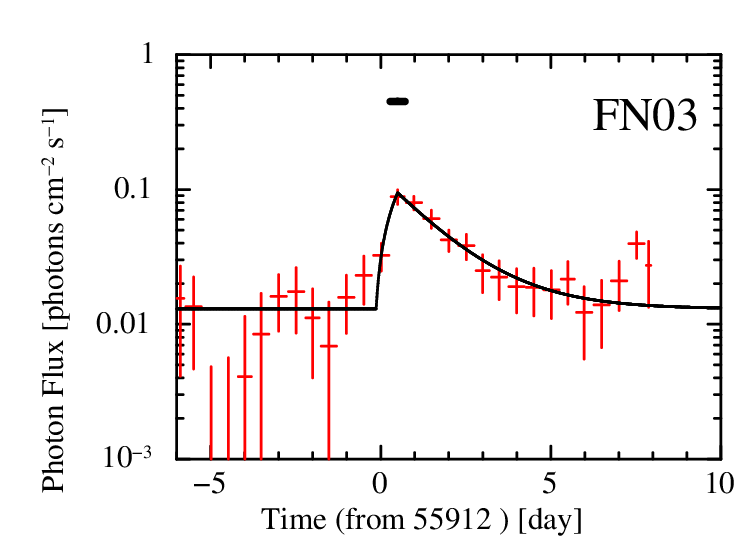}
	\end{center}
	\end{minipage}
	\begin{minipage}{0.5\hsize}
	\begin{center}
		\includegraphics[width=8cm]{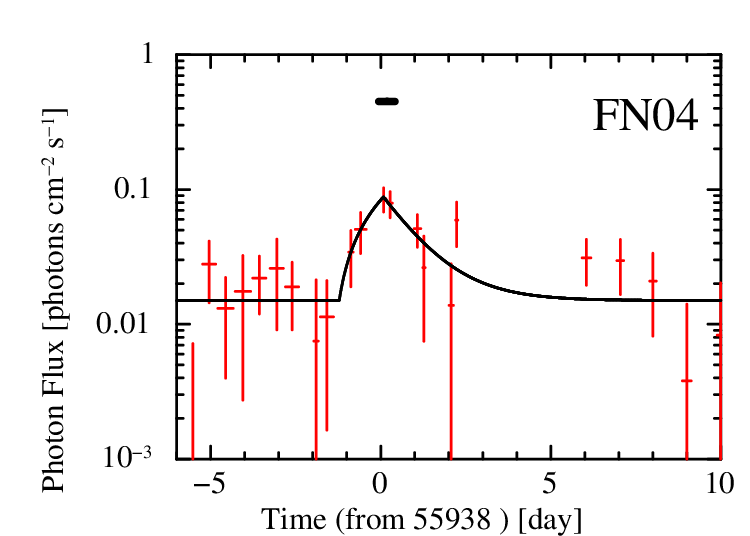}
	\end{center}
	\end{minipage}
	\begin{minipage}{0.5\hsize}
	\begin{center}
		\includegraphics[width=8cm]{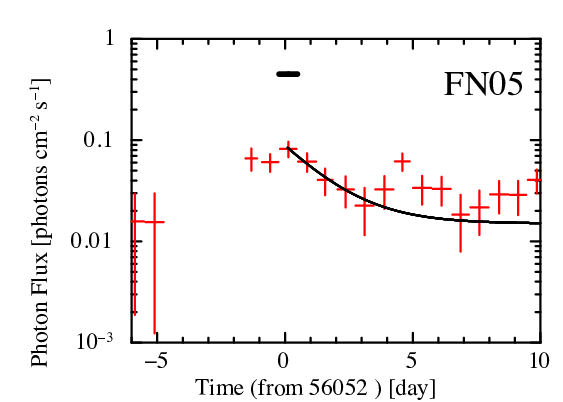}
	\end{center}
	\end{minipage}
	\begin{minipage}{0.5\hsize}
	\begin{center}
		\includegraphics[width=8cm]{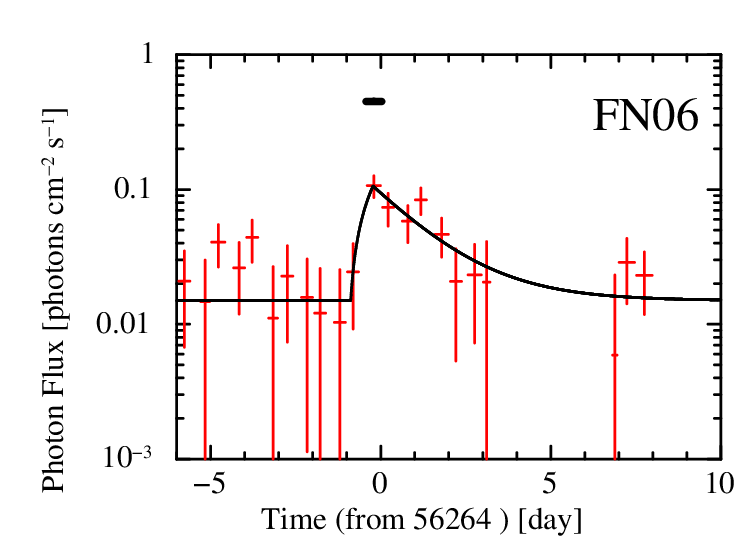}
	\end{center}
	\end{minipage}
	\caption{Light curves of the 11 flares of GT Mus detected with  MAXI in the 2.0--10.0~keV band. The horizontal bar above the peak in each  panel indicates the time interval from which the data are extracted to make the spectrum.}
	\label{fig:flare_lc}
\end{figure*}

\addtocounter{figure}{-1}

\begin{figure*}[t]
	\begin{minipage}{0.5\hsize}
	\begin{center}
		\includegraphics[width=8cm]{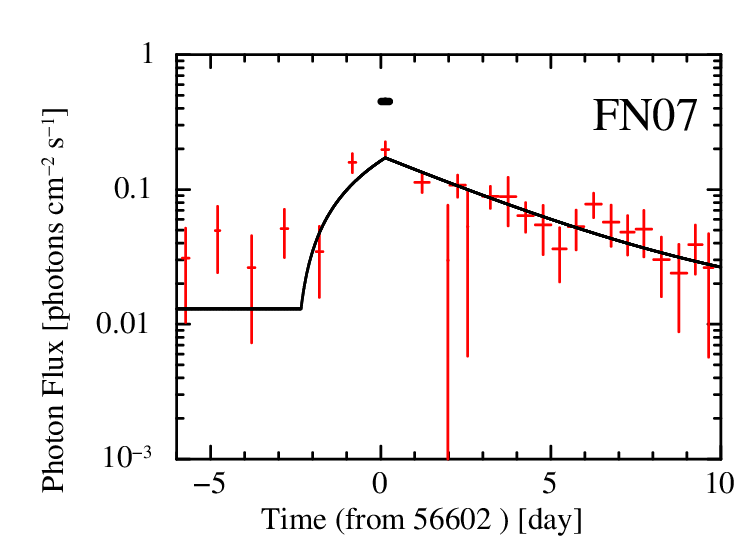}
	\end{center}
	\end{minipage}
	\begin{minipage}{0.5\hsize}
	\begin{center}
		\includegraphics[width=8cm]{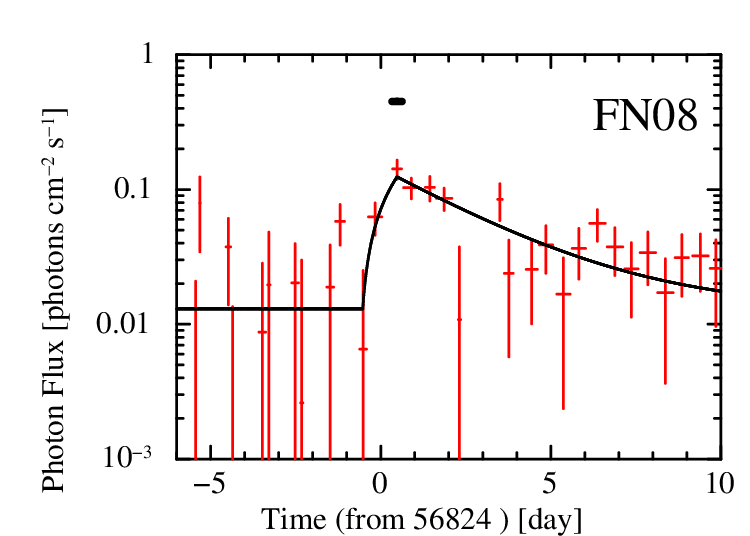}
	\end{center}
	\end{minipage}
	\begin{minipage}{0.5\hsize}
	\begin{center}
		\includegraphics[width=8cm]{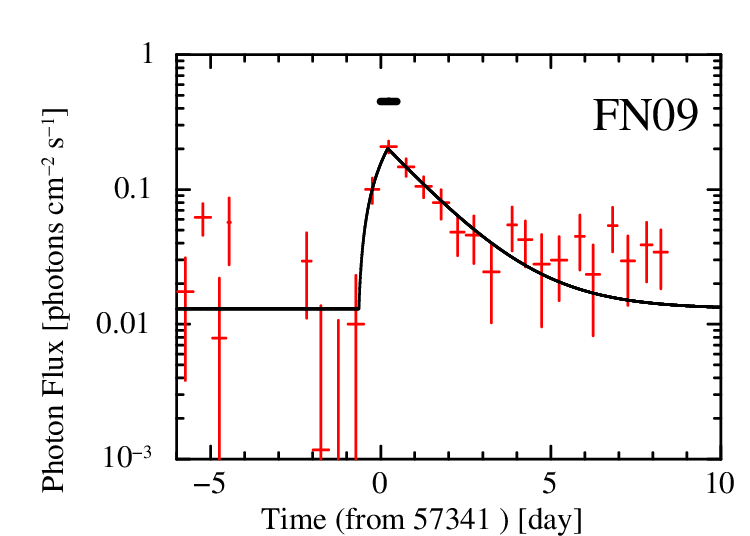}
	\end{center}
	\end{minipage}
	\begin{minipage}{0.5\hsize}
	\begin{center}
		\includegraphics[width=8cm]{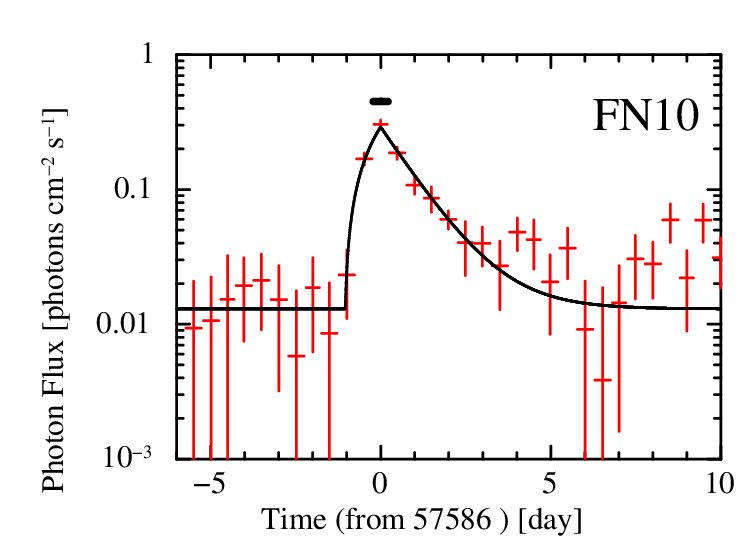}
	\end{center}
	\end{minipage}
	\begin{minipage}{0.5\hsize}
	\begin{center}
		\includegraphics[width=8cm]{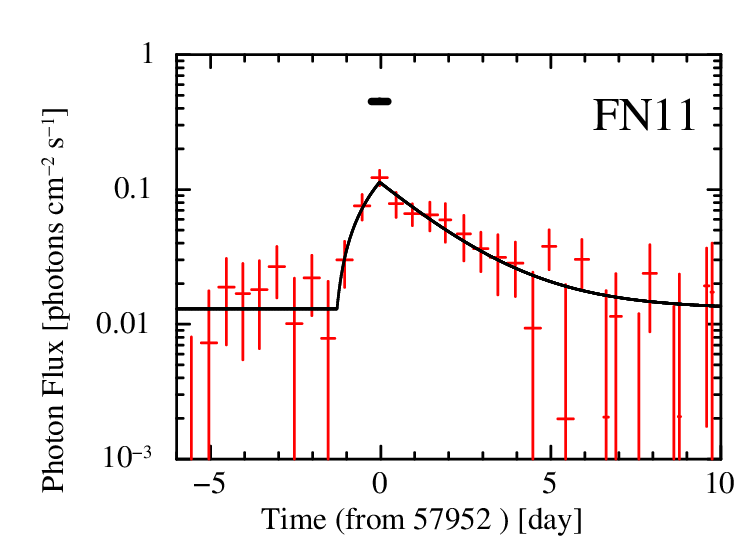}
	\end{center}
	\end{minipage}
	\caption{(continued)}
\end{figure*}

For each of the 11 flares detected with MAXI,
%%MS %confirmed flares with {\maxi},
we performed time series and then spectral analyses. {The duration, or
e-folding time ($\tau_{\rm d}$), of each flare was determined from
the 2.0--10.0~keV light curve with each time bin of half a day
(Figure~$\ref{fig:flare_lc}$ and Table~\ref{tb:maxi_best_fit}).
Each
light curve was fitted with a burst model, which consists of a linear
rise followed by an exponential decay component with an additional constant component.
The constant component was fixed to the
photon flux of the quiescent emission,
0.013~photons~s$^{-1}$~cm$^{-2}$, which was the average value of the over
100~days bin with the Bayesian block process.}

The fitting model was expressed by
  \begin{eqnarray}
   & c(t) & = 0.013 {\rm \  (for\ t \leqq ST)} \nonumber \\
   & c(t) & = {\rm PC} \times \left( \frac{t-{\rm ST}}{{\rm PT}-{\rm ST}} \right) {\rm \  (for\ ST \leqq t \leqq PT)} \nonumber \\
   & c(t) & = {\rm PC} \times exp \left( - \frac{t-{\rm PT}}{\tau_{\rm d}} \right) {\rm \  (for\ PT \leqq t)}  \nonumber .
   \end{eqnarray}
{Here $t$, $c$($t$), ST, PT, and PC are time, count rates, the time when the count rate starts to increase, the time when the count rate is the highest, and the count rate at PT, respectively. Because of the poor statistics, we were not able to determine ST and PT independently. Therefore, we fixed the peak time to the timing of the bin that has the highest photon flux}. The
exceptions are FN~01 and FN~05, whose rising phases were not
observed. {They were fitted instead with an exponential plus the constant
function that describes the quiescent emission. The result showed that
$\tau_{\rm d}$ was 100--360~ks (1--4~day)}.

\begin{center}
	\begin{table*}[htbp]
	\caption{Best-fit parameters of the MAXI light curves and spectra of GT Mus.}
	\begin{center}
	{\scriptsize
	\begin{tabular}{cccccllcc} \hline
		Flare ID & kT & EM & $L_{\rm x,peak}^{a}$ & $\chi_{\rm red}^{2}$(d.o.f.)$^{b}$  & $\tau_{\rm r}$$^{c}$ & $\tau_{\rm d}$ & $E_{\rm rise}$ & $E_{\rm decay}$ \\
		    &   keV   &   10$^{56}$ cm  &   10$^{33}$ \lumcgs &   & ksec & ksec & 10$^{38}$ erg & 10$^{38}$ erg \\ \hline
FN01& 7$^{+27}_{-4}$ & 0.7$^{+0.4}_{-0.2}$ & 1.0$^{+0.2}_{-0.9}$ & 0.47(6) & - & 110$^{+80}_{-50}$ & - & 1.2 \\
FN02& 11$^{+20}_{-5}$ & 1.1$^{+0.3}_{-0.2}$ & 1.9$^{+0.4}_{-1.4}$ & 0.47(7) & 90$^{+10}_{-20}$ & 140$^{+50}_{-40}$ & 0.8 & 2.6 \\
FN03& 6$^{+7}_{-2}$ & 0.7$^{+0.3}_{-0.2}$ & 0.8$^{+0.2}_{-0.6}$ & 1.15(6) & 70$^{+90}_{-20}$ & 110$^{+40}_{-30}$ & 0.3 & 0.9 \\
FN04& 6.6$^{d}$ & 0.7$\pm$0.2 & 0.9$\pm$0.2 & 0.28(6) & 120$^{+40}_{-60}$ & 100$^{+110}_{-50}$ & 0.5 & 0.9 \\
FN05& 5$^{+14}_{-3}$ & 0.9$^{+0.8}_{-0.4}$ & 1.0$^{+0.3}_{-0.9}$ & 0.28(3) & - & 140$^{+180}_{-60}$ & - & 1.3 \\
FN06& 4$^{+7}_{-2}$ & 1.3$^{+1.2}_{-0.6}$ & 1.0$^{+0.2}_{-0.8}$ & 0.08(3) & 60$^{+60}_{-30}$ & 140$^{+100}_{-60}$ & 0.3 & 1.4 \\
FN07& 6.6$^{d}$ & 1.5$\pm$0.4 & 2.0$\pm$0.4 & 0.66(6) & 120$^{+50}_{-40}$ & 360$^{+120}_{-80}$ & 1.1 & 7.3 \\
FN08& 6.6$^{d}$ & 1.0$\pm$0.3 & 1.3$\pm$0.3 & 1.11(5) & 90$^{+20}_{-30}$ & 240$^{+100}_{-80}$ & 0.5 & 3.2 \\
FN09& 6$^{+6}_{-2}$ & 1.7$^{+0.6}_{-0.4}$ & 2.0$^{+0.4}_{-1.0}$ & 1.49(10) & 80$\pm$20 & 130$\pm$40 & 0.7 & 2.7 \\
FN10& 9$^{+7}_{-3}$ & 2.3$^{+0.5}_{-0.4}$ & 3.7$^{+0.4}_{-1.0}$ & 0.59(18) & 90$\pm$10 & 95$\pm$20 & 1.6 & 3.5 \\
FN11& 5$^{+11}_{-2}$ & 1.1$^{+0.6}_{-0.4}$ & 1.2$^{+0.3}_{-0.9}$ & 0.88(5) & 110$^{+20}_{-30}$ & 160$^{+70}_{-50}$ & 0.6 & 1.9 \\
		\hline
	\multicolumn{9}{l}{Errors, upper limits, and lower limits refer to 90\% confidence intervals.} \\
    \multicolumn{9}{l}{$^{a}$ Flare peak luminosity in the 2--20 keV band. {The absorption is corrected.}} \\
    \multicolumn{9}{l}{$^{b}$  $\chi_{\rm red}^{2}$ and  d.o.f. stand for reduced chi-square and degrees of freedom, respectively. Please note that some fittings}\\
    \multicolumn{9}{l}{\ \ \ \ have very low $\chi_{\rm red}^{2}$, primarily due to the low d.o.f., which came from the limited photon statistics.} \\
     \multicolumn{9}{l}{$^{c}$ $\tau_{\rm r}$ is flare rise time, which is difference between flare start time and its peak time.} \\
    \multicolumn{9}{l}{$^{d}$ Because the kT was not derived when we made it free, we fixed to the average value of the other flares.} \\
	\end{tabular} }
	\end{center}
	\label{tb:maxi_best_fit}
	\end{table*}
\end{center}

To determine the physical parameters of the individual flares, we analyzed the spectra at their peaks (see Figure~$\ref{fig:flare_lc}$ for the extracting time regions). In this analysis, we used the optically thin thermal plasma model \texttt{apec} \citep{Smith:2001} to fit the spectra. Given insufficient photon count statistics of the MAXI data, the metal abundance ($Z$) {and the interstellar absorption ($N_{\rm H}$)} in the model {were} fixed at  0.35~$Z_{\sun}$ {and 4.4~$\times$~10$^{20}$~cm$^{-3}$, respectively}, the values derived from the NICER time-resolved spectra  (see Section~\ref{sec:time_res} for details). The redshift was fixed at zero.
The best-fit parameters are shown in Table~$\ref{tb:maxi_best_fit}$.
%\sout{We ignored an interstellar absorption because no significant molecular clouds are known to exist between GT Mus and the Earth and because of the insufficient statistics.}
As a result of the fitting, the absorption-corrected flare peak luminosity in the 2--20~keV band ($L_{\rm X,peak}$),  temperature ($kT$), and  EM were derived to be 1--4 $\times$ 10$^{33}$~\lumcgs, 4--11~keV and {7--23 $\times$ 10$^{55}$~cm$^{-3}$}, respectively. The released flare energies are separately shown, separated for the rise phase and decay phases. The flare energy during the rise phase ($E_{\rm rise}$) was 3--16 $\times$ 10$^{37}$~erg, while that during the decay phase  ($E_{\rm decay}$) was 9--73  $\times$ 10$^{37}$~erg. Then, the emission in the decay phase is a majority of the energy released during the flare.

%The difference between the current version and the original version is that the quiescent emission is subtracted from the data in the current version. However, since the quiescent emission is negligible, no change was seen between the values in original version and the flare energies in decay phase in the current version. A derived flare energy of rise phase ($E_{\rm rise}$) was 3--16 $\times$ 10$^{37}$~erg. That of decay phase  ($E_{\rm decay}$) was 0.9--7.3  $\times$ 10$^{38}$~erg.  Here, we subtracted the quiescent X-ray emission. The $E_{\rm decay}$ is 2--7 times larger than $E_{\rm rise}$.

\subsection{Analysis of the NICER data} \label{sec:time_res}

\subsubsection{Flare phase analysis} \label{sec:nicer_flare}

\begin{figure*}[htbp]
  \begin{center}
  \includegraphics[width=16cm]{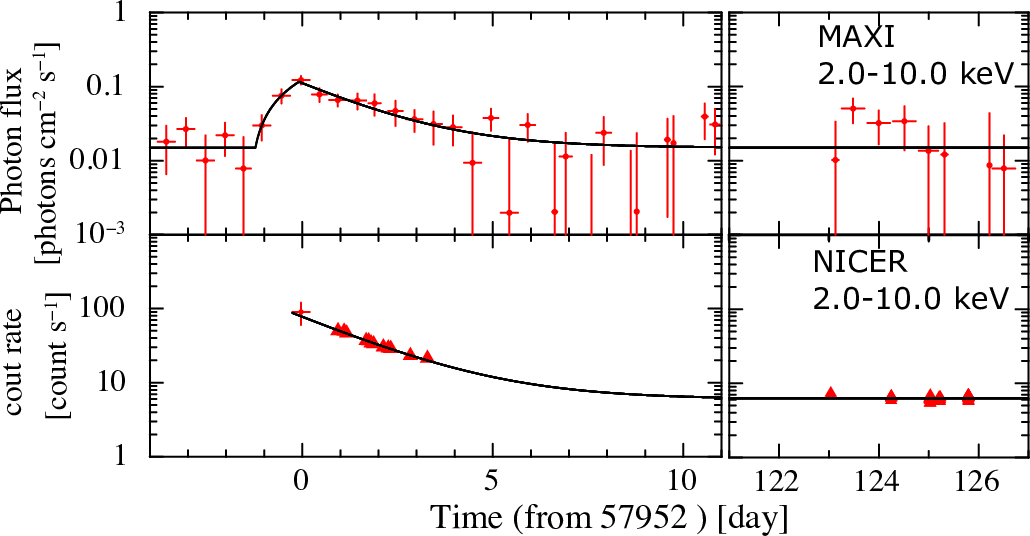}
  \end{center}
\caption{Light curves of FN~11 (see Table~\ref{tb:nicer_best_fit} for time intervals). The upper panel shows the MAXI photon flux. The lower panel shows the NICER count rate, together with that obtained with MAXI (the first bin). The MAXI data was converted to the NICER count rate with {\it Web PIMMS}. The horizontal axis is the number of days since MJD=57,952 (2017 July 18 UT). The solid lines show the fitting function (see text for details).}
  \label{fig:MAXI_NICER_LC}
  \end{figure*}

\begin{figure*}[htbp]
	\begin{minipage}{0.49\hsize}
	\begin{center}
		\includegraphics[width=7.5cm]{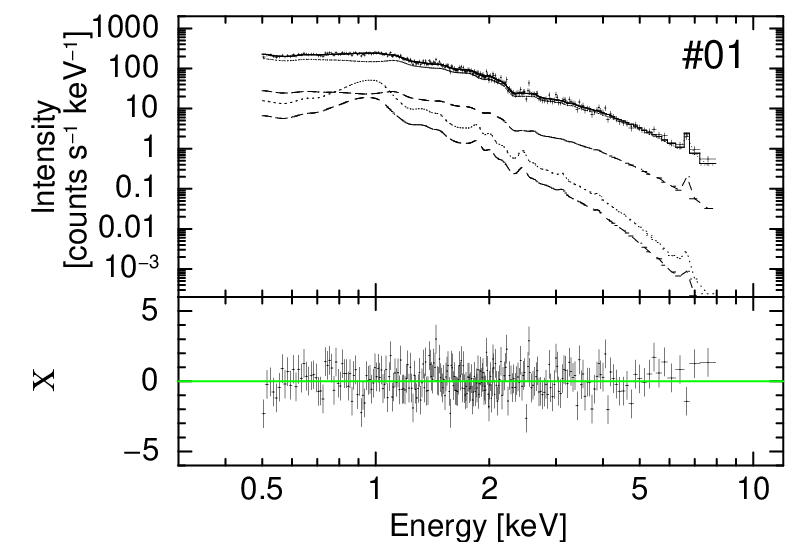}
	\end{center}
	\end{minipage}
	\begin{minipage}{0.49\hsize}
	\begin{center}
		\includegraphics[width=7.5cm]{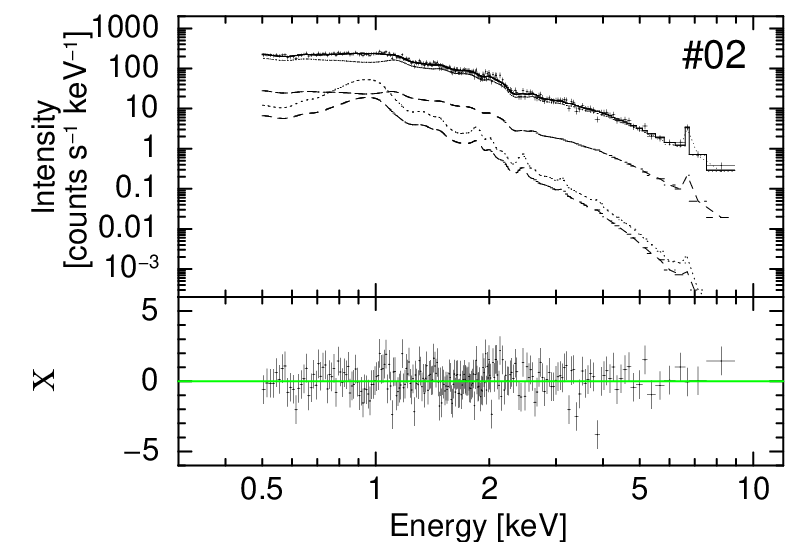}
	\end{center}
	\end{minipage}
	\begin{minipage}{0.49\hsize}
	\begin{center}
		\includegraphics[width=7.5cm]{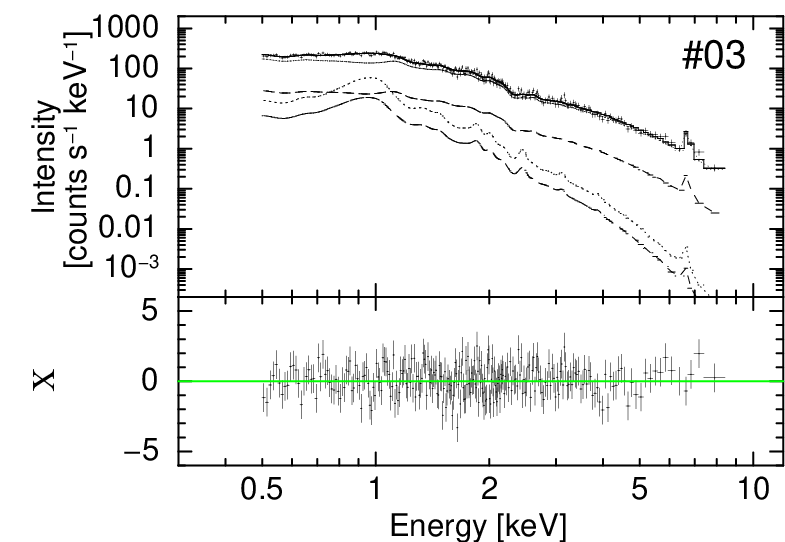}
	\end{center}
	\end{minipage}
	\begin{minipage}{0.49\hsize}
	\begin{center}
		\includegraphics[width=7.5cm]{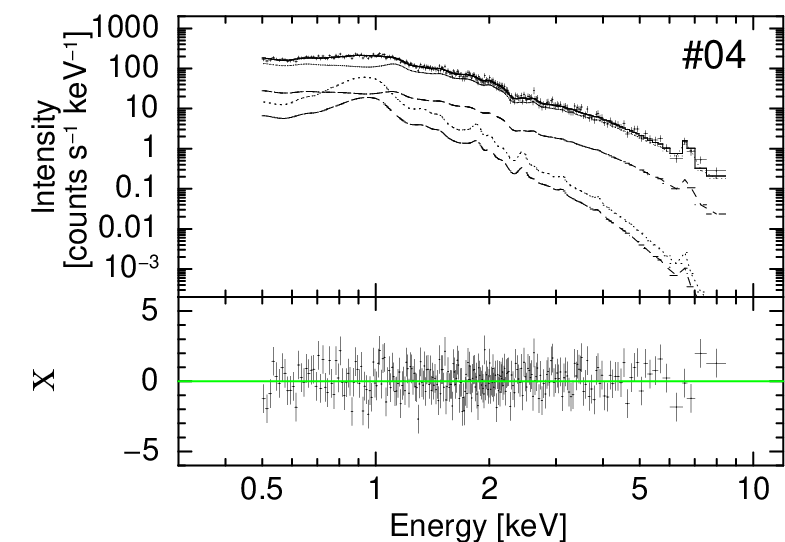}
	\end{center}
	\end{minipage}
	\begin{minipage}{0.49\hsize}
	\begin{center}
		\includegraphics[width=7.5cm]{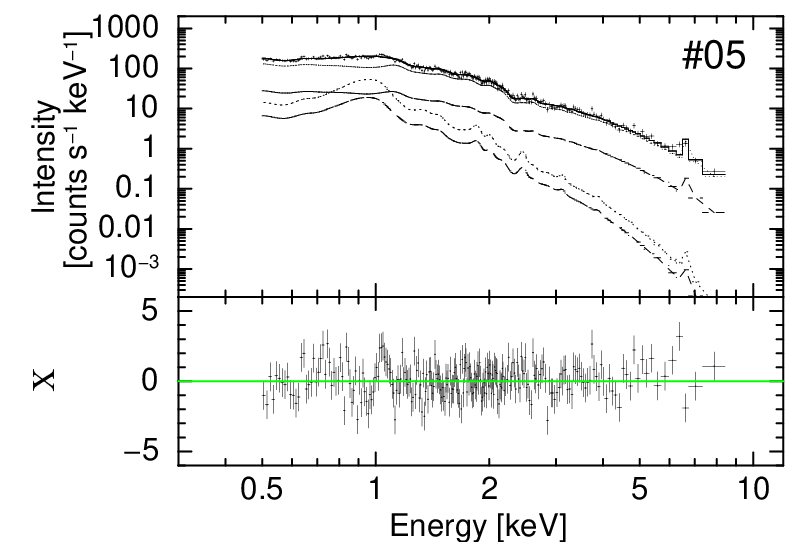}
	\end{center}
	\end{minipage}
	\begin{minipage}{0.49\hsize}
	\begin{center}
		\includegraphics[width=7.5cm]{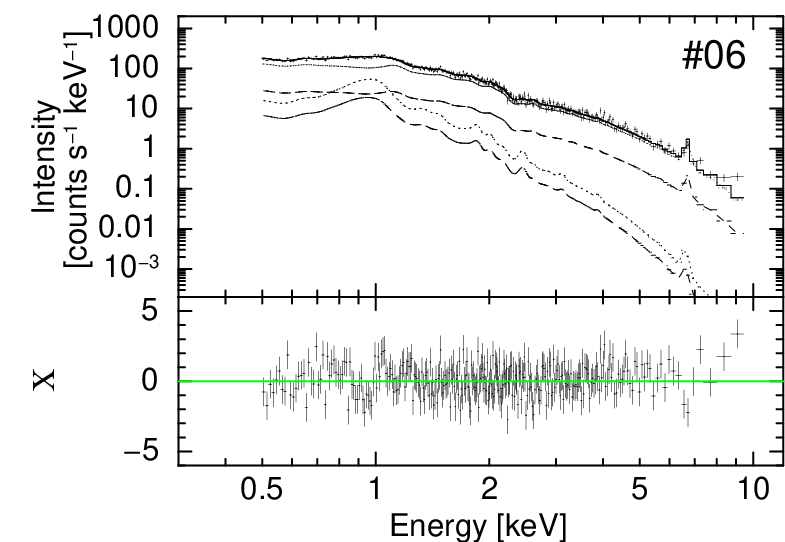}
	\end{center}
	\end{minipage}
	\caption{Time-resolved NICER spectra of  FN~11 (see Table~\ref{tb:nicer_best_fit} for time intervals).  {In each panel, the data and component-separated best-fit model (total model, the flare hot-/cool-temperature components and the quiescent hot-/cool-temperature components shown by solid, dotted, and dashed lines, respectively)} are shown in the upper panel, whereas the $\chi$ values are shown in the lower panel. }
	\label{fig:NICER_SPEC}
\end{figure*}

\addtocounter{figure}{-1}

\begin{figure*}[htbp]
	\begin{minipage}{0.49\hsize}
	\begin{center}
		\includegraphics[width=7.5cm]{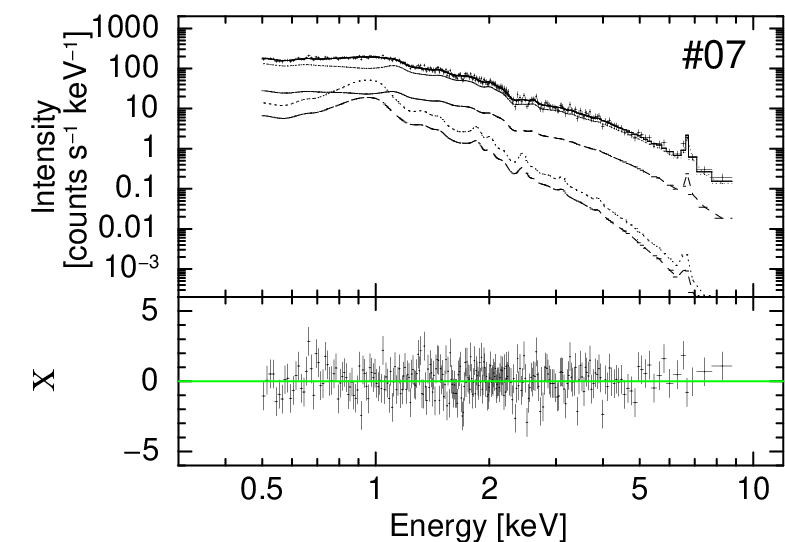}
	\end{center}
	\end{minipage}
	\begin{minipage}{0.49\hsize}
	\begin{center}
		\includegraphics[width=7.5cm]{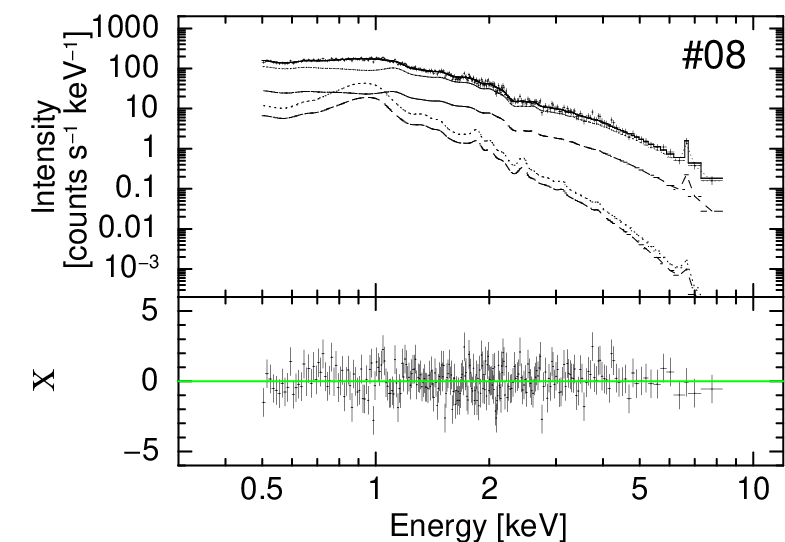}
	\end{center}
	\end{minipage}
	\begin{minipage}{0.49\hsize}
	\begin{center}
		\includegraphics[width=7.5cm]{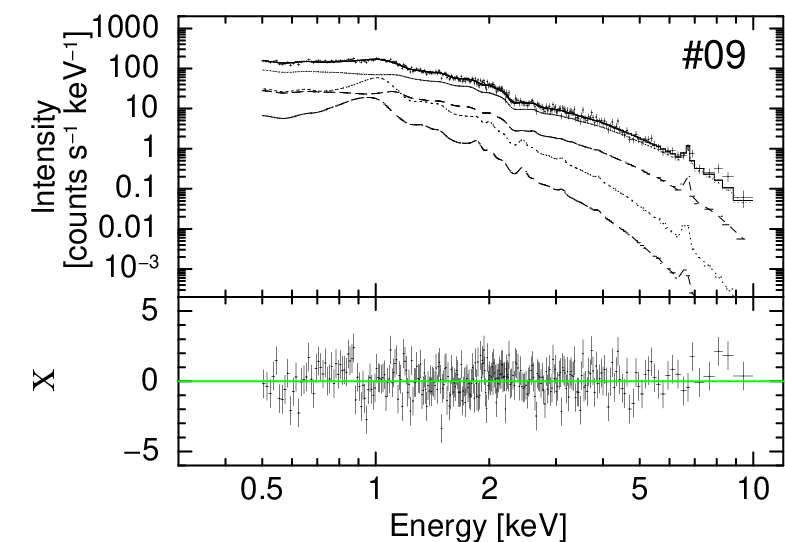}
	\end{center}
	\end{minipage}
	\begin{minipage}{0.49\hsize}
	\begin{center}
		\includegraphics[width=7.5cm]{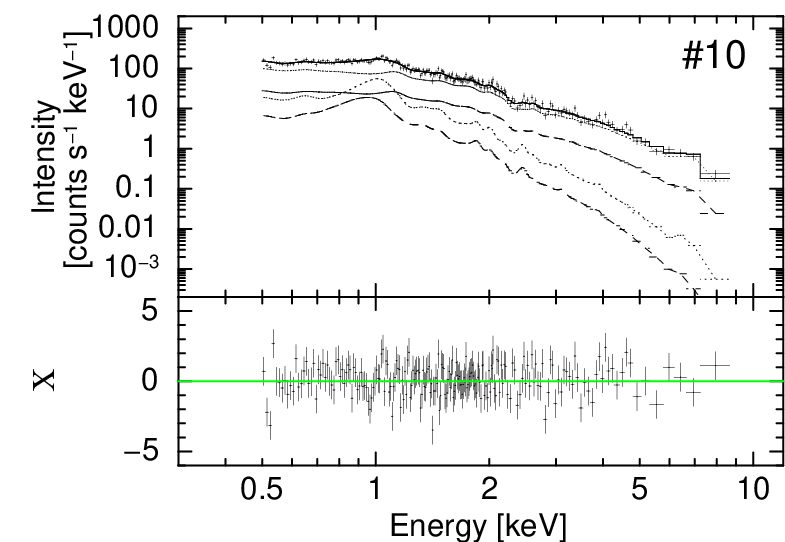}
	\end{center}
	\end{minipage}
	\begin{minipage}{0.49\hsize}
	\begin{center}
		\includegraphics[width=7.5cm]{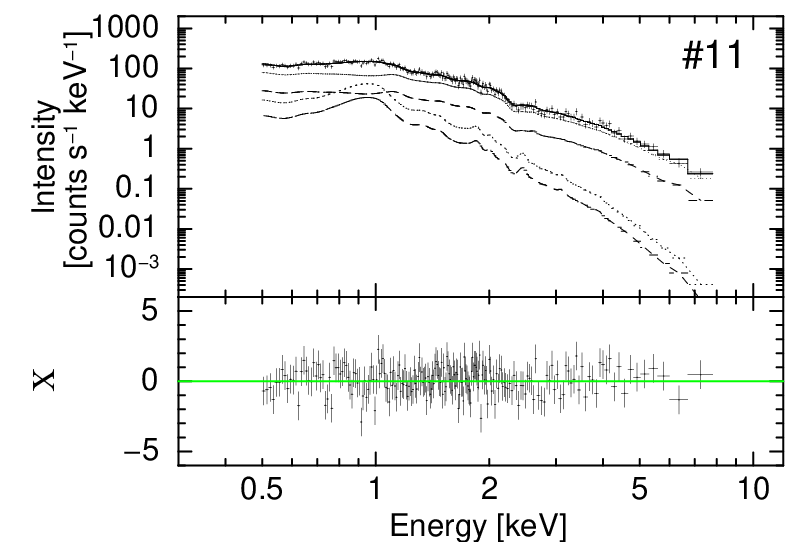}
	\end{center}
	\end{minipage}
	\begin{minipage}{0.49\hsize}
	\begin{center}
		\includegraphics[width=7.5cm]{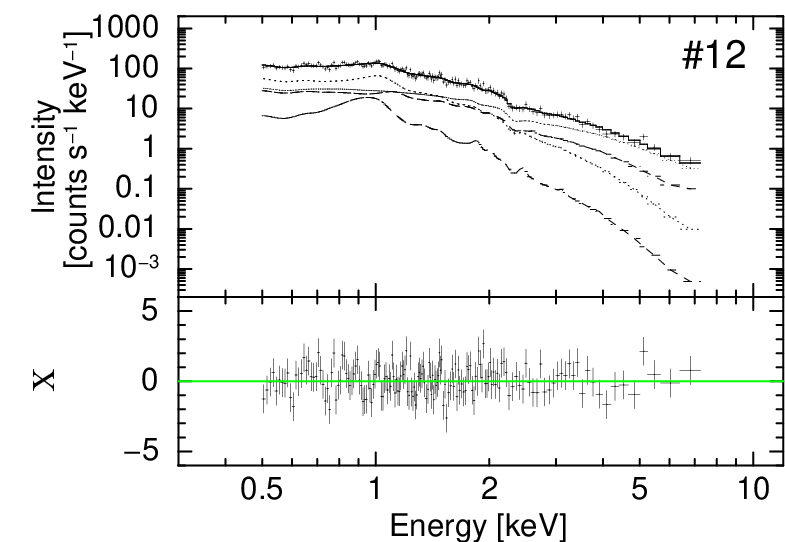}
	\end{center}
	\end{minipage}
	\caption{(continued) }
\end{figure*}

\begin{table*}[htbp]
	\begin{rotatetable*}
	\caption{Best-fit parameters for the NICER time-resolved spectra during FN11}
	{\scriptsize
	\renewcommand\arraystretch{1}
	\begin{tabular}{ccccccccccc} \hline
		 No.$^{a}$ & Time Interval$^{b}$ & $N_{\rm H}^{c}$ & $Z^{d}$ & $kT_{\rm hot}$$^{e}$ & EM$_{\rm hot}$$^{f}$ & $kT_{\rm cool}$$^{e}$ & EM$_{\rm cool}$$^{f}$ & Count Rate  & $L_{\rm X}^{g}$ & $\chi_{\rm red}^{2}$(d.o.f.)$^{h}$\\
		    &   days &   10$^{20}$~cm$^{3}$  &   $Z_{\sun}$  &   keV &   10$^{54}$~cm$^{-3}$ &   keV &   10$^{54}$~cm$^{-3}$ &   counts~sec$^{-1}$  &   10$^{32}$~\lumcgs   &   \\ \hline
01 & T0+0.840:T0+1.034 & 5$\pm$1 & 0.3$\pm$0.1 & 4.5$^{+0.4}_{-0.3}$ & 55$\pm$2 & 1.05$^{+0.26}_{-0.06}$ & 4$^{+8}_{-1}$ & 292$\pm$2 & 9.4$\pm$0.1 & 0.92(240) \\
02 & T0+1.097:T0+1.098 & 4$\pm$1 & 0.5$\pm$0.1 & 4.8$\pm$0.4 & 49$\pm$2 & 1.01$^{+0.06}_{-0.08}$ & 3$\pm$1 & 287$\pm$2 & 9.3$^{+0.1}_{-0.2}$ & 0.86(219) \\
03 & T0+1.162:T0+1.163 & 4$\pm$1 & 0.4$\pm$0.1 & 4.5$\pm$0.3 & 48$\pm$2 & 1.05$^{+0.16}_{-0.05}$ & 4$\pm$1 & 280$\pm$2 & 8.8$\pm$0.1 & 1.07(248) \\
04 & T0+1.676:T0+1.678 & 5$\pm$1 & 0.4$\pm$0.1 & 3.9$\pm$0.3 & 38$\pm$2 & 0.99$\pm$0.05 & 4$\pm$1 & 236$\pm$1 & 7.1$\pm$0.1 & 0.96(250) \\
05 & T0+1.741:T0+1.743 & 5$\pm$1 & 0.4$\pm$0.1 & 4.1$^{+0.3}_{-0.2}$ & 38$\pm$1 & 1.03$\pm$0.04 & 4$\pm$1 & 228$\pm$1 & 6.9$\pm$0.1 & 1.25(239) \\
06 & T0+1.805:T0+1.807 & 4$\pm$1 & 0.4$\pm$0.1 & 3.8$\pm$0.2 & 37$\pm$1 & 1.04$^{+0.03}_{-0.04}$ & 4$\pm$1 & 224$\pm$1 & 6.57$^{+0.04}_{-0.07}$ & 1.16(294) \\
07 & T0+1.870:T0+1.872 & 3$\pm$1 & 0.4$\pm$0.1 & 4.0$\pm$0.2 & 34$\pm$1 & 1.03$\pm$0.04 & 3$\pm$1 & 219$\pm$1 & 6.5$\pm$0.1 & 0.98(260) \\
08 & T0+2.128:T0+2.130 & 5$\pm$1 & 0.3$\pm$0.1 & 3.8$\pm$0.2 & 32$\pm$2 & 0.99$^{+0.05}_{-0.06}$ & 3$\pm$1 & 197$\pm$1 & 5.9$\pm$0.1 & 0.98(249) \\
09 & T0+2.256:T0+2.259 & 3$^{+2}_{-1}$ & 0.2$^{+0.1}_{-0.2}$ & 5.1$^{+5.3}_{-0.6}$ & 25.5$^{+1.9}_{-7.2}$ & 1.27$^{+0.07}_{-0.10}$ & 7$^{+20}_{-3}$ & 186$\pm$1 & 5.6$\pm$0.1 & 1.03(276) \\
10 & T0+2.320:T0+2.321 & 2$\pm$1 & 0.5$\pm$0.2 & 4.7$^{+1.0}_{-0.7}$ & 25$\pm$2 & 1.26$^{+0.09}_{-0.22}$ & 4$^{+5}_{-2}$ & 185$\pm$2 & 5.4$\pm$0.1 & 1.21(203) \\
11 & T0+2.837:T0+2.838 & 6$^{+4}_{-1}$ & 0.2$^{+0.1}_{-0.2}$ & 3.6$^{+2.5}_{-0.4}$ & 24.9$^{+2.0}_{-8.1}$ & 1.03$^{+0.33}_{-0.08}$ & 5$^{+8}_{-2}$ & 163$\pm$1 & 4.7$\pm$0.1 & 0.89(201) \\
12 & T0+3.284:T0+3.285 & 5$^{+2}_{-4}$ & $<$0.15 & 4$<$ & 12.5$^{+7.0}_{-1.9}$ & 1.29$^{+0.07}_{-0.27}$ & 16$^{+9}_{-14}$ & 144$\pm$1 & 4.4$^{+0.1}_{-1.0}$ & 1.01(161) \\
\hline
\multicolumn{10}{l}{Errors refer to 90\% confidence intervals.} \\
\multicolumn{10}{l}{$^{a}$ Serial number of each time interval.} \\
\multicolumn{10}{l}{$^{b}$ The time interval is in units of days. T0 = 57,952 [MJD] (2017 July 18 UT 00:00:00).} \\
\multicolumn{10}{l}{$^{c}$ Hydrogen column density.} \\
\multicolumn{10}{l}{$^{d}$ Metal abundance.} \\
\multicolumn{10}{l}{$^{e}$ Plasma temperature ($kT$) of the hot and cool components.} \\
\multicolumn{10}{l}{$^{f}$ The EM of the hot and cool components.} \\
\multicolumn{10}{l}{$^{g}$ The X-ray luminosity in the 0.5--10.0 keV band. {The absorption is corrected.}} \\
\multicolumn{10}{l}{$^{h}$ Here $\chi_{\rm red}^{2}$ and  d.o.f.  stand for reduced chi-square and degrees of freedom, respectively.} \\
	\end{tabular} }
	\label{tb:nicer_best_fit}
	\end{rotatetable*}
\end{table*}

\begin{figure*}[htbp]
    \begin{center}
    \includegraphics[width=16cm]{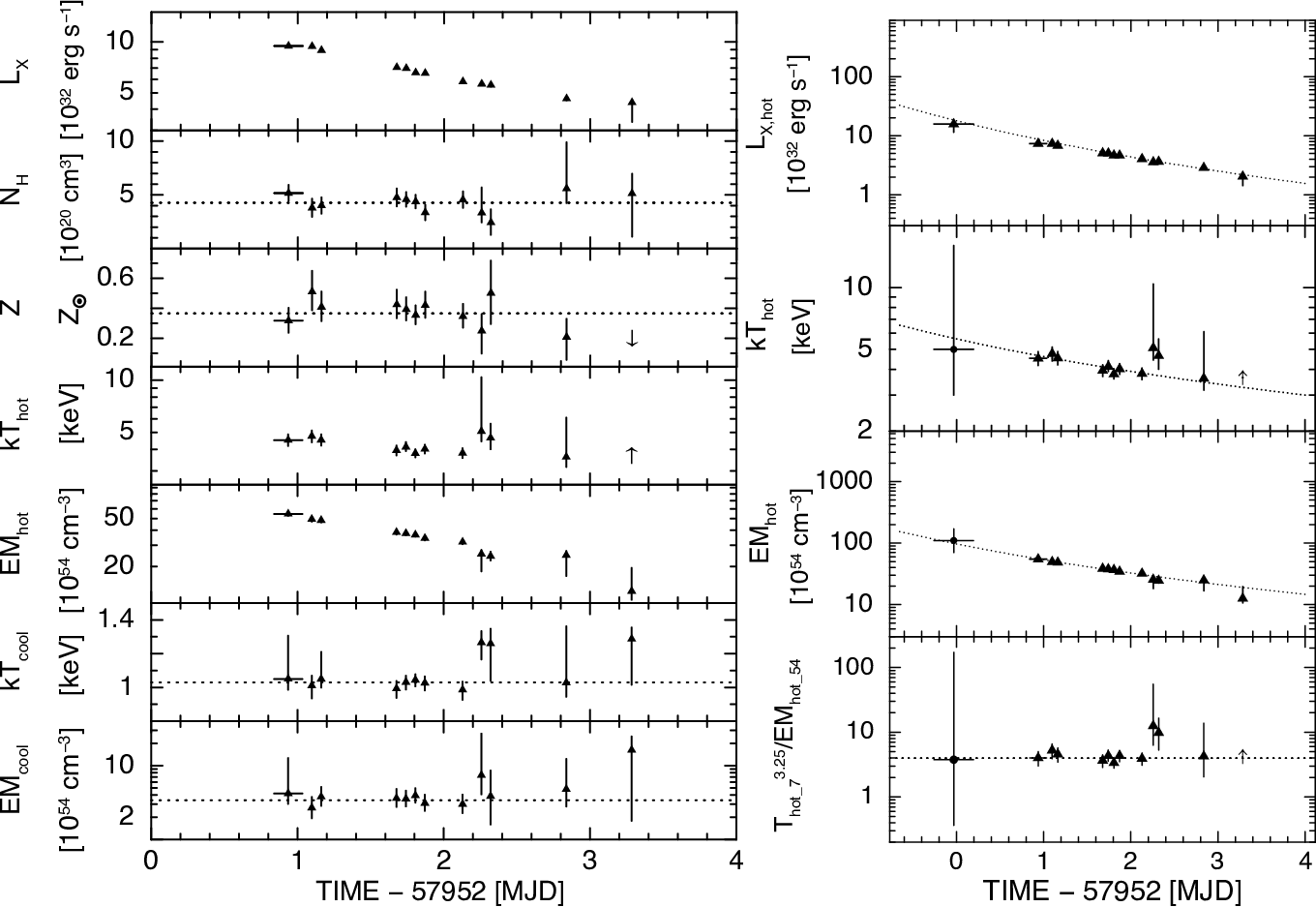}
    \end{center}
    \caption{{(Left) From top to bottom, the time variations of the luminosity in the 0.5--10~keV band in units of 10$^{32}$~\lumcgs, hydrogen column density in 10$^{20}$~cm$^{3}$, metal abundance compared to solar abundance, temperatures of the hot and cool plasma components in keV, and their EMs in 10$^{54}$~cm$^{-3}$ are shown for FN~11 from the NICER time-resolved spectra (Table~\ref{tb:nicer_best_fit}). Dotted lines are the  best-fit models summarized in Table~\ref{tb:const_fit}. (Right) Time variation of the parameters obtained for the hot component. The first bin is from MAXI data, while the other bins are from NICER data. From the top, the luminosity in the 0.5--10~keV band, the temperature, the EM, and the ratio of $T_{\rm hot}^{3.25}$/EM$_{\rm hot}$ are shown. Dotted lines in the panels for $T_{\rm hot}$ and EM$_{\rm hot}$ show the best-fit model for Equation~\ref{eq:kT_EM_fitting}. The dotted line in the luminosity plot is calculated  from the models for $T_{\rm hot}$ and EM$_{\rm hot}$. In all the panels, the NICER and MAXI data are shown by triangles and circles, respectively. }}
    \label{fig:NICER_TIME_RESOLVED}
\end{figure*}

NICER follow-up observations were performed  for FN~11. The MAXI and NICER light curves in the 2.0--10.0 keV band of this flare are shown in Figure~\ref{fig:MAXI_NICER_LC}. {Note that since the flare peak was missed with NICER, the MAXI peak data were added into the Figure~\ref{fig:MAXI_NICER_LC}, after conversion to the NICER count rate with {\it Web PIMMS} \footnote{\url{https://heasarc.gsfc.nasa.gov/cgi-bin/Tools/w3pimms/w3pimms.pl}}.}
The NICER light curve was fitted with an exponential function with the decay constant $\tau_{\rm d}$ and a constant function, the latter of which was fixed to the NICER count rate in the quiescent state of 6.3~count~s$^{-1}$ in the 2.0--10.0~keV band. The period for the quiescent state is 3 days from 2017 November 18 to 2017 November 20 (see Section~\ref{subsec:nicer}).
As a result, the decay constant  $\tau_{\rm d}$ was derived to be 174$\pm$3~ks. The reduced $\chi^{2}$ ($\chi_{\rm red}^{2}$) and degrees of freedom (d.o.f.) were 1.8 and {10}, respectively. The derived $\tau_{\rm d}$ is consistent with the value derived from the MAXI light curve.

%The latter %%MS %The constant component
%was fixed to the count rate of the quiescent state observed with NICER. At first, we derived the quiescent state count rate. We fitted the count rate in the 2.0--10.0~keV band , which is obtained from 2017--11--18 to 2017--11--20, with a constant function fitting (the right lower panel of Figure~\ref{fig:MAXI_NICER_LC}). As a result, we obtained 6.3 count per second. In this fitting, the reduced chi-squared ($\chi_{\rm red}^{2}$) and degree of freedom (d.o.f.) were 1.33 and 56, respectively. Because no time variation was confirmed, we considered that these data show the quiescent state X-ray emission.
%%MS %to
%the NICER  count rate in the quiescent state (see Section~\ref{subsec:nicer}), which is 6.0 counts per second in the 2--10~keV band.
%Finally, using the quiescent state count rate, we fitted the flare light curve. The model (an exponential plus a constant) well reproduced the light curve. The estimated e-folding time, $\tau_{\rm d} = 174 \pm 3$~ks, is consistent with that derived from the MAXI light curve in the same flare. In this fitting, the $\chi_{\rm red}^{2}$ and d.o.f. were 1.8 and 9, respectively.
%%MS %Consequently, the e-folding time $\tau_{\rm d}$ was derived to be 174$\pm$3~ks. The reduced chi-square value was 1.8. The derived $\tau_{\rm d}$  is consistent with the one derived  from the \maxi\ light curve.

\begin{center}
	\begin{table}[htbp]
	%\begin{minipage}
	\caption{ {Best-fit parameters of the stable parameters in Table~\ref{tb:nicer_best_fit} with a constant model}}
	%\centering
	\begin{center}
	{\scriptsize
	\renewcommand\arraystretch{1}
	\begin{tabular}{ccccc} \hline
        &   $N_{\rm H}$    &   $Z$ &   $kT_{\rm cool}$  &   EM$_{\rm cool}$ \\
        &   10$^{20}$ cm$^3$    &   $Z_{\sun}$  &   keV &   10$^{54}$~cm$^{-3}$ \\ \hline
    C$^{a}$  & 4.2$\pm0.3$ & 0.37$\pm0.03$ & 1.03$_{-0.02}^{+0.01}$ & 3.4$\pm0.4$ \\
    $\chi_{\rm red}^{2}$~$^{b}$  & 0.7 & 0.6 & 0.8 & 0.3 \\ \hline
    \multicolumn{5}{l}{Errors refer to 90\% confidence intervals.} \\
    \multicolumn{5}{l}{The d.o.f. of all the fitting is 11.} \\
    \multicolumn{5}{l}{$^{a}$ Fitting result of time series with a constant function.} \\
    \multicolumn{5}{l}{$^{b}$ Reduced chi-squared.} \\
	\end{tabular} }
	\end{center}
	\label{tb:const_fit}
	%\end{minipage}
	\end{table}
	\end{center}

We performed time-resolved spectroscopy using the NICER data divided into 12 time intervals, which  correspond to  12 ISS orbits, in the 0.5--10~keV band listed in Table~\ref{tb:nicer_best_fit}. Figure~\ref{fig:NICER_SPEC} shows all of the time-resolved spectra and the best-fit model. {Since the spectra are the sum of the quiescent and flare emissions, the modeling of the quiescent emission was fixed to the best-fit values given in Section~\ref{sec:nicer_qui}.}
{For the flare component, we first adopted an absorbed one-temperature optically thin thermal plasma model. Here we used \texttt{TBabs} \citep{Wilms:2000} and \texttt{apec} for the absorption and the thin thermal plasma models, respectively. We also fixed} the red-shift to zero. However, this model was rejected with $\chi_{\rm red}^{2}>$2. Then, we fitted the flare component with an absorbed two-temperature optically thin thermal plasma model with the metal abundances ($Z$) of the cool and hot plasma components  assumed to be the same. The spectra were found to be well reproduced by this model. Table~\ref{tb:nicer_best_fit} and Figure~\ref{fig:NICER_TIME_RESOLVED} summarize the best-fit parameters and their time variations, respectively.

To check the potential time variations of these NICER spectral parameters, we fitted the time series of the hydrogen column density ($N_{\rm H}$), $Z$, the $kT$ of the cool component ($kT_{\rm cool}$), and the EM of the cool component (EM$_{\rm cool}$) with a constant model and found that all of these parameters were constant from flare to flare within the statistics. Table~\ref{tb:const_fit} shows the fitting results of the time series of the four parameters {summarized in Table~\ref{tb:nicer_best_fit}}. By contrast, the parameters EM and $kT$ of the hot component (EM$_{\rm hot}$ and $kT_{\rm hot}$) showed clear declining trends. We study them in detail in Section~\ref{subsec:coolpro}.

\subsubsection{Quiescent state analysis} \label{sec:nicer_qui}

\begin{figure}[htbp]
  \begin{center}
  \includegraphics[width=8cm]{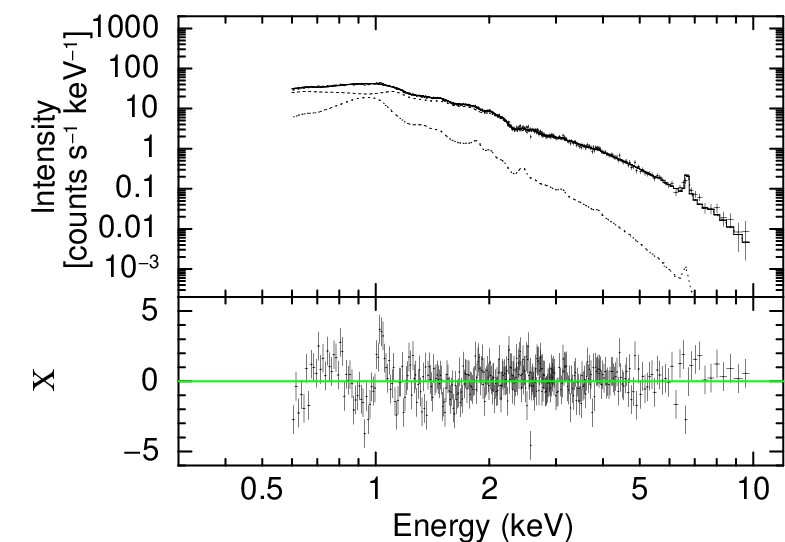}
  \end{center}
\caption{Time-averaged quiescent-state spectra. The data and component-separated best-fit model (the total model and its hot-/cool-temperature components are shown with solid and dashed lines, respectively) are shown in the upper panel, whereas the $\chi$ values are shown in the lower panel.}
  \label{fig:quiescent_spectrum}
  \end{figure}

We analyzed a quiescent-state spectrum as follows,  using the time-averaged spectra for 3 days (from 2017 November 18 to 2017 November 20). The total exposure was 5~ks. The spectrum is shown in Figure~\ref{fig:quiescent_spectrum}. %Thanks to the long exposure time, the statistic is significantly higher than the each time interval of the follow-up observation.
The spectrum could be fitted with
%\sout{the same model as in the flare-spectral analysis described in the previous subsection.}
{the absorbed two thin thermal plasma models.} As a result, the $kT_{\rm hot,q}$, EM$_{\rm hot,q}$, $kT_{\rm cool,q}$, and EM$_{\rm cool,q}$ values, where the subscript ``${\rm q}$'' means the quiescent state, were determined to be 3.13$^{+0.06}_{-0.07}$ keV, 8.7$^{+0.02}_{-0.01} \times 10^{54}$ cm$^{-3}$, 1.02$\pm0.01$~keV and 1.98$\pm0.02 \times 10^{54}$~cm$^{-3}$, respectively. The parameters $N_{\rm H,q}$ and $Z_{\rm q}$ were $5.9\pm0.3 \times 10^{20}$~cm$^{3}$ and 0.24$^{+0.01}_{-0.02}$~$Z_{\sun}$, respectively. {The  absorption-corrected quiescent X-ray luminosity in the 0.5--10.0~keV band was 1.09$\pm$0.01~$\times$~10$^{32}$~\lumcgs.} The resultant $\chi_{\rm red}^{2}$ and d.o.f. were 1.30 and 308, respectively. {The cool component in the quiescent state can be intrinsically interpreted as the same as the cool component during the flare, having a similar temperature and EM (see Table~\ref{tb:const_fit}).}
%\sout{The derived $N_{\rm H,q}$, $Z_{\rm q}$, $kT_{\rm cool,q}$ and $EM_{\rm cool,q}$ are almost consistent with the constant fitting result (Table~\ref{tb:const_fit})}.

\section{Discussion}\label{sec:dis}

\subsection{Flare Parameters of GT Mus }\label{subsec:energy}

\begin{figure}[htbp]
  \begin{center}
  \includegraphics[width=8cm]{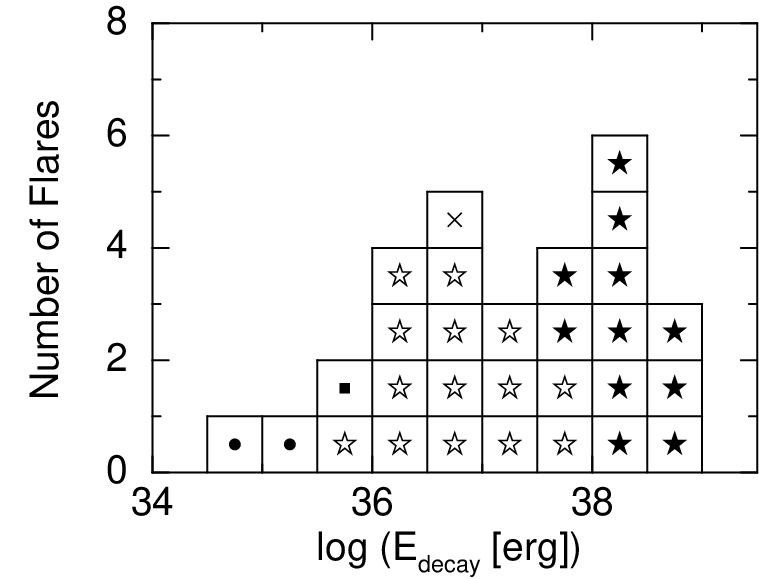}
  \end{center}
\caption{ Flare energies of GT Mus compared with those from other stars reported in \cite{Tsuboi:2016}. Filled stars, open stars, filled circles, filled squares, and crosses in boxes indicate GT Mus, RS CVn-type stars (except for GT Mus), dMe stars, Algol, and TWA 7, respectively. The flare energies of the other flares, extracted from \citet{Tsuboi:2016}, were recalculated using distances of the Gaia DR2 catalog \citep{Gaia:2016, Gaia:2018} }
  \label{fig:Ehisto}
  \end{figure}

  \begin{figure}[htbp]
  \begin{center}
  \includegraphics[width=8.5cm]{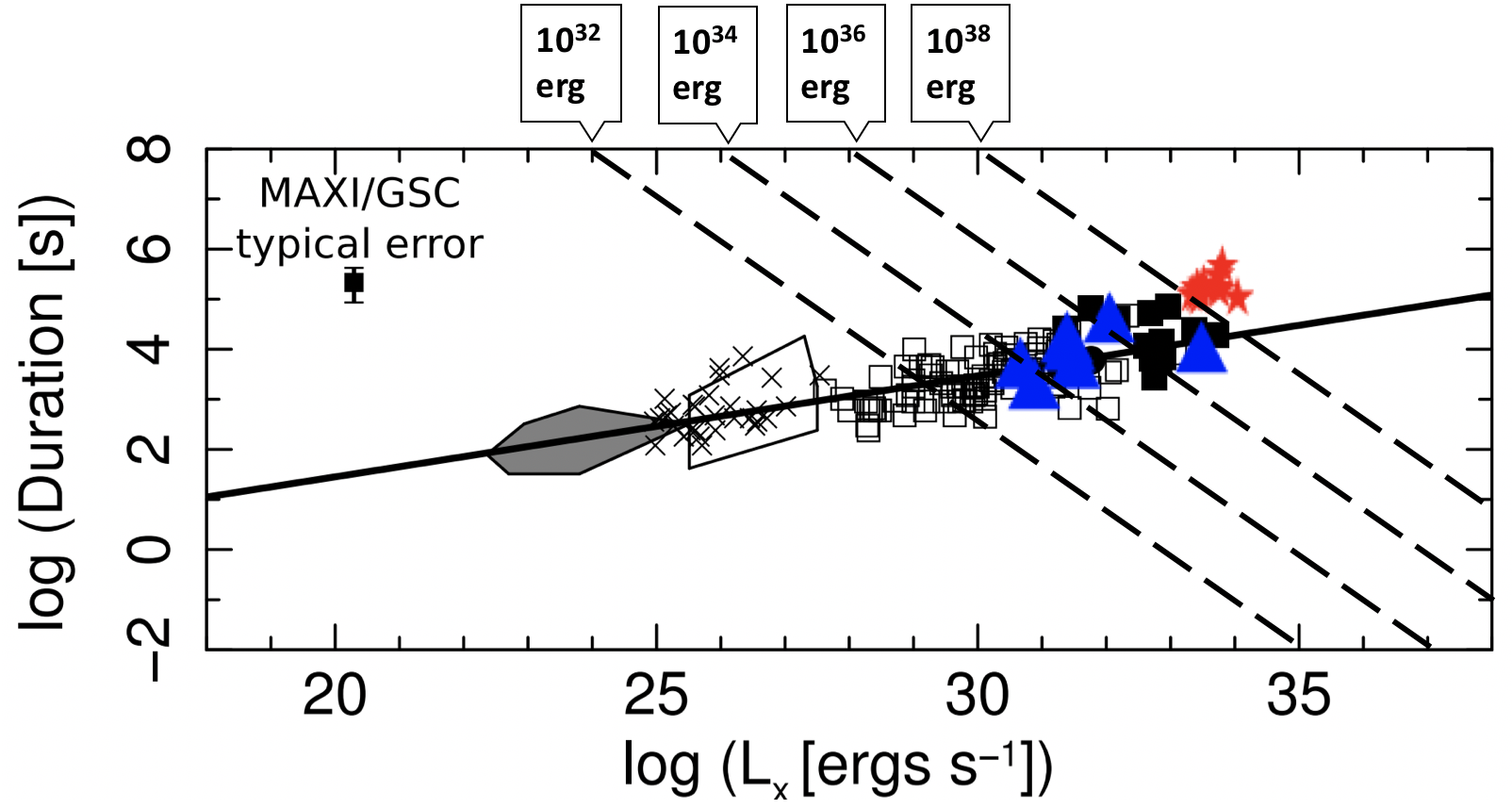}
  \end{center}
\caption{{The universal correlation of the duration of flares vs. X-ray luminosity obtained by \cite{Tsuboi:2016}, which is established from solar flares to large stellar flares. The blue triangles are RS CVn type flares \citep{Pandey:2012, Nordon:2007, Drake:2014, Gong:2016, Osten:2007}. The red stars are GT Mus flares from this work. The dashed line show total released flare energies of 10$^{32}$--10$^{38}$~erg.}}
  \label{fig:Energy_contour}
  \end{figure}

Using MAXI, we have detected 11 flares from GT Mus. {From the X-ray spectroscopy, GT Mus was found to have large EMs and high plasma temperatures when it is flaring. Both of the parameters are confirmed to be located at the upper end of the universal EM--$kT$ correlation when we plot them into the Figure~4 of \citet{Tsuboi:2016}.
}

{The large EMs and the plasma temperatures give the large X-ray luminosities. The intrinsic X-ray luminosities in the 0.1--100 keV band, derived with the procedure described in the appendix of \citealt{Tsuboi:2016}, are in the range of 2--5~$\times$~10$^{33}$~\lumcgs. Our timing analysis showed long durations ($\tau_{\rm r} + \tau_{\rm d}$) of 2--6 days with long decay times ($\tau_{\rm d}$) of 1--4 days. We plotted both parameters on the $\tau_{\rm d}$--$L_{\rm X}$ diagram of \citet{Tsuboi:2016} (their Figure~5), adding the samples of RS CVn binaries detected with XMM-Newton \citep{Pandey:2012}, Chandra \citep{Nordon:2007, Drake:2014, Gong:2016}, and Swift \citep{Osten:2007} after converting their luminosity ranges to the 0.1--100~keV band by using their temperatures and EMs. The results are shown in Figure~\ref{fig:Energy_contour}.
In the diagram, the GT Mus flares are located at the upper end of the universal correlation obtained by \citet{Tsuboi:2016}.}

{The large X-ray bolometric luminosities and the decay time scales give the large X-ray released energies during the flares. In Figure~\ref{fig:Energy_contour}, the dashed lines indicate the same energies. The energy released in the GT Mus flare decay phase is in the range 1--11~$\times$~10$^{38}$~erg in the 0.1--100 keV band, and that for the whole duration ($\tau_{\rm r} + \tau_{\rm d}$) is in the range 2--13~$\times$~10$^{38}$~erg. They are about more than an order of magnitude larger than the other observed stellar flares. We further show the observed flare energy distribution obtained with MAXI (this work and \citealt{Tsuboi:2016}) in Figure~\ref{fig:Ehisto}. Here also, the extremely large energies of the GT Mus flares are shown. }

{All these results indicate that the GT Mus flares are among the hottest, longest, and brightest flares ever observed.}

%{When we consider the other stellar flare, not included in these correlations, they are consistent with the stellar flares other than MAXI ($kT$:2.4--13~keV, log EM:53--56, log $L_{\rm X,bol}$:31--33.5, log $\tau_{\rm d}$:3.2--4.5) \citep{Pandey2012,Gong:2016,Osten:2007}} Therefore, the GT Mus flares are among the hottest, longest, and brightest flares ever observed.

\subsection{High flare activity}\label{subsec:activity}

During the MAXI 8-year observation period, the large flares were detected every year.
Although the possibility that GT Mus is always active is not excluded,  it is likely that GT Mus was in an active phase over 8 yr.

It is well known that the Sun has an 11-years sunspot cycle \citep{Schwabe:1844}. Its X-class flares ($\sim$10$^{31}$~erg) tend to occur in the period near the solar maximum, spanning about a half of a solar cycle ($\sim$5.5~years) \citep{Aschwanden:2012}.
On the other hand, activity cycles of other RS CVn-type stars have been obtained from a spot number/area as 14--20 years for HR~1099 (V711~Tau) \citep{Lanza:2006, Muneer:2010, Perdelwitz:2018}, 9.2 years from II Peg \citep{Lindborg:2013}, and 15 years from LQ Hya \citep{Berdyugina:2002}. {During the half of these activity cycles, the stars are active, which is indicative from the existence many/large spots.}

If the same trend of the activity-phase ratio applies to GT Mus, the activity cycle of GT Mus would be at least sixteen years long.
Future monitoring observations will determine how long the active phase of GT Mus lasts, if it indeed has a cycle like other active stars.

\subsection{Cooling process of the GT Mus flares}\label{subsec:coolpro}

{We investigate the cooling of the hot component of FN~11, which is the dominant flaring event in the observed flares. Here we apply the quasi-static cooling model of \citet{vandenOord_Mewe:1989} as a simple approximation. }

In the  model, the  ratio between the radiative-cooling timescale ($\tau_{\rm r}$) and the conductive-cooling timescale ($\tau_{\rm c}$)  remains constant during  the flare decay phase. The two timescales can be expressed by the following formulae:
    \begin{equation}
    \tau_{\rm r} = \frac{3 n_{\rm e} kT}{n_{\rm e}^{2} \Psi(T)}
    \label{eq:trad}
    \end{equation}
    \begin{equation}
    \tau_{\rm c} = \frac{3 n_{\rm e} kT}{E_{\rm c}},
    \label{eq:tcond}
    \end{equation}
where $T$, $n_{\rm e}$, $\Psi(T)$, and $E_{\rm c}$ are the temperature, electron density, emissivity of an optically thin thermal plasma, and mean conductive energy loss rate, respectively. Here $\Psi(T)$ in  Equation~\ref{eq:trad} is a combination of X-ray emission lines and bremsstrahlung continuum spectrum. In the case of the hot component of FN~11, it is given by $\Psi(T)$ = $\Psi(T)_{0}~T^{-{\gamma}}$ = $10^{-24.73}~T^{1/4}$ erg cm$^{3}$ s$^{-1}$, since the flare plasma temperature is  higher than 20~MK \citep[e.g.][]{Mewe:1985, Mewe:1986} throughout the observation.

We assume that the flare geometry is one semicircular loop having a constant cross section, as \cite{vandenOord_Mewe:1989} did. Under this assumption, $E_{\rm c}$ in Equation~\ref{eq:tcond} is expressed by
    \begin{equation}
    E_{\rm c} = \frac{16 \kappa_{0} T^{7/2}}{7 L^{2}},
    \label{eq:cond_loss}
    \end{equation}
where $\kappa_0$ and $L$ are  the plasma thermal conductivity of 8.8 $\times$ $10^{-7}$~erg~cm$^{-1}$~s$^{-1}$~K$^{-7/2}$ \citep{Spitzer:1962} and  the flare loop full length, respectively. In this case, from the hydrodynamic equations of conservation of mass, momentum, and energies, the solution yields the scaling law in \cite{Kuin:1982} or Equations~19b, 19c, and 20 in \cite{vandenOord_Mewe:1989}, and then the ratio $\tau_{\rm r}$/$\tau_{\rm c}$ of 0.1.

When we assume that the aspect ratio ($a$ : diameter-to-length ratio) and $L$ do not change during a flare, the fact that  $\tau_{\rm r} / \tau_{\rm c}$ is not time-variable means that $T^{13/4}/EM$ is not time-variable either, from Equations (1)--(3) in this work \citep[they are corresponding to Equations 7, 8, 9 and 10 in][]{vandenOord_Mewe:1989}.
The bottom right panel of Figure~\ref{fig:NICER_TIME_RESOLVED} shows the value of $T_{\rm hot\_7}^{13/4}$/EM$_{\rm hot\_54}$ as a function of time, where $T_{\rm hot\_7}$ and EM$_{\rm hot\_54}$ are $T_{\rm hot}$/(10$^{7}$ K) and EM$_{\rm hot}$/(10$^{54}$~cm$^{-3}$), respectively. We apply a constant function to this plot and obtain an acceptable fit with $T_{\rm hot\_7}^{3.25}$/EM$_{\rm hot\_54}$~=~2.9$\pm0.3$, a $\chi_{\rm red}^{2}$ value of 0.8, and a d.o.f. of 11. Then the FN~11 flare can be described with the model.

\subsection{Flare Loop Geometry}\label{subsec:loop_geo}

In the case of quasi-static flare cooling, the decay timescale of $kT_{\rm hot}$ and EM$_{\rm hot}$ can be estimated on the basis of a set of time-evolution formulae expressed in the form of
    \begin{equation}
    F(t) = F(t') \left ( 1 + \frac{t - t'}{3\tau_{\rm qs}} \right )^{-\alpha}
    \label{eq:kT_EM_fitting}
    \end{equation}
according to Equation~26 and 27 in \cite{vandenOord_Mewe:1989}, where $F$ represents either $kT_{hot}$ or EM$_{hot}$ as a function of time ($t$). Here $\tau_{\rm qs}$ is called the quasi-static time scale, the specific timescale that determines the decay of both $kT_{hot}$ and EM$_{hot}$.
The value of $\alpha$ depends on $F$: 8/7 or 26/7 for the cases where $F$ is $kT_{\rm hot}$  and EM$_{\rm hot}$, respectively \citep{vandenOord_Mewe:1989}.

We fit
%\sout{our}
{the time variation of $kT_{\rm hot}$ and EM$_{\rm hot}$} obtained with {MAXI and} NICER
%\sout{data}
with Equation~\ref{eq:kT_EM_fitting} simultaneously, with the common parameter $\tau_{\rm qs}$ free. The parameter $t'$ is set to -0.032 (MJD=57,951.968), which corresponds to the epoch at which the peak of $kT_{\rm hot}$ and EM$_{\rm hot}$ were observed with MAXI. {Then we obtained $\tau_{\rm qs}=$130$\pm$4~ksec, $kT_{\rm hot}(t')=$5.7$\pm$0.2~keV, $EM_{\rm hot}(t')$=99$\pm$2 with $\chi_{\rm red}^{2}$/d.o.f.=0.89/22. The best-fit models are shown in the time variation of $kT_{\rm hot}$ and EM$_{\rm hot}$ in the right panel of Figure~\ref{fig:NICER_TIME_RESOLVED}, while the model for $L_{\rm X, hot}$ is calculated from those for $kT_{\rm hot}$ and EM$_{\rm hot}$, and inserted in the upper right panel. }
%\sout{The $F(t)$, i.e.,  $kT_{\rm hot}(t)$ and $EM_{\rm hot}(t)$ to estimate cooling timescales (referred to as $\tau_{\rm qs,kT}$ and $\tau_{\rm qs,EM}$, respectively)}.
%\sout{In this formulation, the two timescales ($\tau_{\rm qs,kT}$ and $\tau_{\rm qs,EM}$) are expected to be the same when the flare cools quasi-statically (see the previous subsection), and we confirm it is indeed the case within the errors; $\tau_{\rm qs,kT}$ $\approx$ $\tau_{\rm qs,EM}$. In the following discussion, we  use  $\tau_{\rm qs,EM}$  as the cooling timescale of FN~11, for it is better constrained than $\tau_{\rm qs}$}.

We estimate from these values the three geometric parameters of FN~11 of flare loop length $L$, aspect ratio of the loop $a$, and electron density $n_{\rm e}$ using the following equations of  the quasi-static cooling model  \citep{vandenOord_Mewe:1989, Tsuboi:2000},
   \begin{eqnarray}
    a = & & 1.38 \left( \frac{\tau_{\rm qs}}{10~{\rm ks}}\right)^{-1/2} \nonumber \\
    & &\left\{ \frac{kT_{\rm hot}(t')}{\rm keV} \right \} ^{-33/16} \left\{ \frac{EM_{\rm hot}(t')}{10^{54}~{\rm cm}^{-3}} \right \}^{1/2},
    \end{eqnarray}
    \begin{equation}
    L = 1.0R_{\sun} \frac{\tau_{\rm qs}}{10~{\rm ks}} \left\{ \frac{kT_{\rm hot}(t')}{\rm keV} \right\}^{7/8},
    \end{equation}
    \begin{equation}
    n_{\rm e} = 4.4 \times 10^{10} {\rm cm}^{-3} \left( \frac{\tau_{\rm qs}}{10~{\rm ks}} \right)^{-1} \left \{ \frac{kT_{\rm hot}(t')}{\rm keV} \right \}^{3/4},
    \end{equation}
and obtain the following:
{
   \begin{eqnarray}
    a & = & 0.11\pm0.02 \\
    L & = & (4.2\pm0.2) \times 10^{12}\ \ [\mathrm{cm}]\ \ (61\pm5 R_{\sun})\\
    n_{\rm e} & = & (1.3\pm0.1) \times 10^{10}\ \ [\mathrm{cm^{-3}}].
    \end{eqnarray}
 }

The estimated loop length $L$ is much larger, by two orders of magnitude, than that of the typical solar flare, 10$^{9}$--10$^{10}$~cm \citep{Kontar:2011}. {\cite{Pandey2012} made a comprehensive loop length comparisons of RS CVn-type stars using pointed observations (e.g. XMM-Newton). They found loop lengths of 10$^{10}$--10$^{12}$~cm. On the other hand,} %\sout{According to}
{MAXI has observed large flares from RS CVn type stars that have a loop length of 10$^{11}$--10$^{13}$~cm \citep{Tsuboi:2016}. The derived loop length in this work is among the highest in the MAXI flare sample. The derived GT Mus loop length}
%\sout{the derived loop}
is then almost four times larger than the stellar radius, 16.56~$R_{\sun}$ (1.2~$\times$~10$^{12}$~cm) \citep{Gaia:2016, Gaia:2018}. There are the MAXI/GSC sources that have flares with loop lengths of up to an order of magnitude larger than the stellar radius; our derived ratio of $L$ relative to the stellar radius of GT Mus ranges within the nominal range for these other MAXI stellar flares. Note that the binary separation of HD~101379 is unknown; hence, we  are unable to  tell whether the loop is connected between the RS CVn-type stars.

The estimated parameter, $a$, is  within the range for solar active-region loops \citep[0.06--0.2;][]{Golub:1980}. The footprints of the loop cover {$\sim$3.7\%} of the stellar surface (6.7~$\times~10^{23}$~cm$^{2}$).
The estimated density $n_{\rm e}$ is
%\sout{almost}
consistent with the typical solar and stellar flares of 10$^{10}$--10$^{13}$~cm$^{-3}$ \citep{Aschwanden:1997, Gudel:2004, Reale:2007}.

%Finally, we calculate the ratio $\tau_{\rm r}/\tau_{\rm c}$ to be $\sim$0.1, substituting the obtained plasma parameters to Equations~\ref{eq:trad} and \ref{eq:tcond}.   This means that 90\% of the thermal energy was lost by radiation. It is consistent with the theoretical prediction by \cite{vandenOord_Mewe:1989} for quasi-static cooling.

\subsection{Coronal magnetic activity}\label{subsec:quiescent}
\begin{figure}[htbp]
  \begin{center}
  \includegraphics[width=8.5cm]{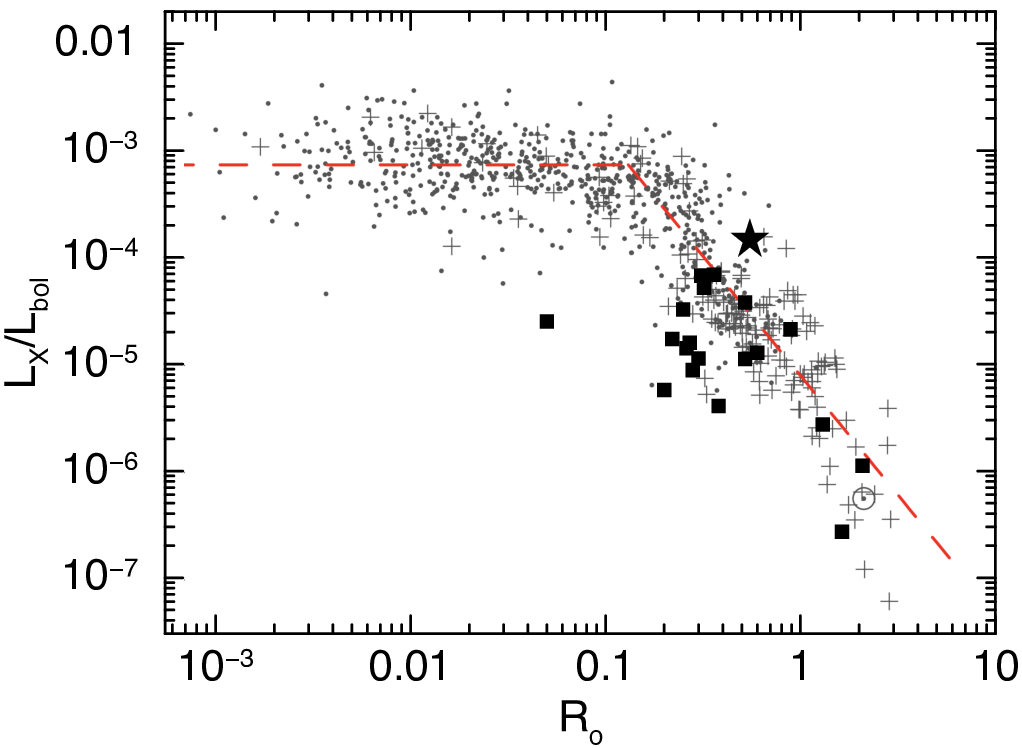}
  \end{center}
\caption{Scatter plot of the X-ray to bolometric luminosity ratio ($L_{\rm X}$/$L_{\rm bol}$) vs. Rossby number ($R_{\rm o}$). Dots and plus signs are for late-type main-sequence single and binary stars, respectively. The solar symbol  is for the Sun \citep{Wright:2011}. Squares  are for G- and K-type giant binaries \citep{Gondoin:2007}. The star indicates GT Mus.}
  \label{fig:rosby_number}
  \end{figure}

In this section, we examine the magnetic activity of GT Mus. As for low-mass ($<$1.5$M_{\sun}$) main-sequence stars, one of the indicators of magnetic activity, the X-ray to bolometric luminosity ratio ($L_{\rm X}$/$L_{\rm bol}$), is well known to show a good correlation to the Rossby number ($R_{\rm o}$), which is the ratio of the rotation period to the convective turnover timescale \citep[e.g. ][]{Wright:2011}. Whereas the data show a distinctive log-linear relation between $L_{\rm X}$/$L_{\rm bol}$ and $R_{\rm o}$ for $R_{\rm o}$~$\gtrsim$~0.1, the relationship is flat for $R_{\rm o}$~$\lesssim$~0.1. This flattening indicates saturation of magnetic activity. In contrast, as for the intermediate-mass giant binaries like GT Mus, which has the mass of $M_{\ast}$~=~2.7$M_{\sun}$ \citep{Tokovinin:2008}, the relation has not yet been established, though some studies exist for a period-activity relationship using the other parameters \citep[e.g. ][]{Gondoin:2007}.

We replotted the data points of 20 G- and K-type intermediate-mass giant binaries in \cite{Gondoin:2007}, changing the vertical axis of the surface X-ray flux to $L_{\rm X}$/$L_{\rm bol}$, and keeping the independent variable as $R_{\rm o}$ (see their Figure~2, right panel). Here $R_{\rm o}$ and $L_{\rm bol}$ are used from \cite{Gondoin:2007}, calculated with the stellar parameters taken from the literature \citep{Schrijver:1991, Strassmeier:1993, McDowell:1994, Hummel:1994, Voges:1999, Kovri:2001, Williamon:2005}. Each $L_{\rm X}$ is taken from the ROSAT bright source catalog \citep{Voges:1999} in order to unify the X-ray band with that used in \cite{Wright:2011}, 0.1--2.4 keV. The distribution of data points for the giant binaries is found to agree with the relation derived from that of  late-type main-sequence stars (see Figure~\ref{fig:rosby_number}).

We then evaluated GT Mus in the diagram. The value $R_{\rm o}$ of GT Mus is derived to be 0.614 from the rotation period of 61.4~days \citep{Murdoch:1995} and the convective turnover timescale of 100~days, the latter of which was obtained from
a function of effective temperature during the evolution of a 2.2~$M_{\sun}$ star \citep{Gunn:1998}. Here the effective temperature (4761~K) is taken from the \cite{Gaia:2016, Gaia:2018}. The $L_{\rm bol}$ was calculated to be 4.9~$\times$~10$^{35}$~erg from the effective temperature and stellar radius of 16.56~$R_{\sun}$. With extrapolation of the quiescent-state spectrum of NICER, the value of $L_{\rm X}$ is estimated to be 7.18~$\pm 0.02$~$\times$~10$^{32}$\lumcgs\ in the 0.1--2.4~keV band. Figure~\ref{fig:rosby_number} compares the location of GT Mus with other stars in the $L_{\rm X}$/$L_{\rm bol}$-$R_{\rm o}$ diagram.

We find that GT Mus is consistent with the trend followed by late-type main-sequence stars and G- and K-type giant binaries.  However, we note that GT Mus shows considerably higher $L_{\rm X}$/$L_{\rm bol}$ ratio than the other giant binaries in the diagram. This high X-ray fraction supports the idea that GT Mus is in an active phase.

\section{Summary} \label{sec:sum}

\begin{enumerate}

\item
MAXI detected 11 flares from the RS CVn type star GT Mus in its 8 yr of all-sky X-ray monitoring observations. The released energies during the decay phases of the flares were in the range of {1--11~$\times$~10$^{38}$~erg} in the 0.1--100 keV band, which is higher than any other stellar flares detected in 2 yr of monitoring observations with MAXI \citep{Tsuboi:2016} as well as the other flares detected with the other missions. The released energies during whole duration ($\tau_{\rm r} + \tau_{\rm d}$) ranged {2--13~$\times$~10$^{38}$~erg} in the same band. The flare parameters ($kT$, EM, X-ray luminosity in the 0.1--100 keV band, and $\tau_{\rm d}$) are found to be located at the upper end of the known parameter correlation plot of stellar flares compiled by \cite{Tsuboi:2016}, suggesting that these flares have the largest energy ever observed from stellar flares.

\item
We performed a 3 day follow-up X-ray observation of GT Mus with NICER from 2017 July 18, 1.5~days after the MAXI detection of a large flare. The time-resolved spectra suggest that the flare cooled quasi-statically during the NICER observation. On the basis of a quasi-static cooling model, the flare loop size is estimated to be {4.2$\pm$0.2 $\times$ 10$^{12}$~cm (61$\pm5$~$R_{\sun}$)}. This size is a 2--3 orders of magnitude larger than that of the typical solar flare loop of 10$^{9}$--10$^{10}$~cm.

\item
For the first time, we plotted the G and K giant binary samples in the diagram of X-ray to bolometric luminosity ratio versus Rossby number and obtained a consistent distribution with those for the low-mass stars. The Rossby number and log($L_{\rm X}$/$L_{\rm bol}$) of GT Mus are 0.614 and  $-$3.5, respectively, which puts GT Mus in line with the relation derived from low-mass and giant binary stars in the diagram. It shows a considerably higher $L_{\rm X}$/$L_{\rm bol}$ than other giant binaries. This high X-ray fraction suggests that GT Mus is at a high magnetic activity level, which is consistent with what is inferred from its recurring large flares.

%It would be an interesting and unresolved question whether a flare mechanism in the  GT Mus system is different from that in the other stars.
%%MS %The parameter of GT Mus of
%The X-ray to bolometric luminosity ratio against the Rossby number \citep{Durney:1978} ($L_{\rm X}$/$L_{\rm bol}$ vs. $R_{\rm o}$) is found to be consistent with  the known distribution
%%MS %of the parameter of
%for low-mass main sequence stars.
%The $L_{\rm X}$/$L_{\rm bol}$ value of GT Mus is estimated to be $-$3.5, which is %%MS %located at the upper end of the scale,
%at a highest level in low-mass main sequence and intermediate-mass giant stars,
%and hence is consistent with the observed very high activity of GT Mus. Among the giant binaries, GT Mus shows high $L_{\rm X}$/$L_{\rm bol}$. This high X-ray fraction suggests that GT Mus is at a high magnetic activity level, consistent with what is inferred from its recurring large flares.

\end{enumerate}

\section*{Acknowledgments}
This research used MAXI data provided by RIKEN, JAXA, and the MAXI team. This work was supported by NASA through the NICER team and the Astrophysics Explorers Program. It  used data and software provided by the High Energy Astrophysics Science Archive Research Center (HEASARC), a service of the Astrophysics Science Division at NASA/GSFC and the High Energy Astrophysics Division of the Smithsonian Astrophysical Observatory. The authors thank Kazunari Shibata for useful discussions and comments on the cooling process. R.S. acknowledges financial support  from the Junior Research Associate Program in RIKEN, JSPS Overseas Challenge Program for Young Researchers and Research Assistant Program in Chuo University. Y.T. acknowledges financial support  from JSPS KAKENHI grant Number JP17K05392. W.I. acknowledges support  from the Special Postdoctoral Researchers Program in RIKEN and JSPS KAKENHI grant Number JP16K17717. Y.M. gratefully acknowledges funding from the Tanaka Kikinzoku Memorial Foundation. M.S. acknowledges financial support from JSPS KAKENHI grant No. JP16K17672.

\bibliography{gtmus_bibliography}

%% This command is needed to show the entire author+affilation list when
%% the collaboration and author truncation commands are used.  It has to
%% go at the end of the manuscript.
%\allauthors

%% Include this line if you are using the \added, \replaced, \deleted
%% commands to see a summary list of all changes at the end of the article.
%\listofchanges

\end{document}